\documentclass[apj,twocolappendix,appendixfloats]{emulateapj}
\usepackage{graphicx}
\usepackage{pslatex}
\usepackage{color}
\usepackage{natbib}
\usepackage{amsmath}
\usepackage{times}
\usepackage{epstopdf}
\usepackage{soul}

\def\lesssim{\lower.5ex\hbox{$\; \buildrel < \over \sim \;$}}
\def\gtrsim{\lower.5ex\hbox{$\; \buildrel > \over \sim \;$}}
\newcommand{\msun}{\mbox{${M}_\odot$}}

\newcommand{\fesc}{\mbox{$f_{\rm esc}$}}
\newcommand{\fescg}{\mbox{$\left<f_{\rm esc}\right>$}}
\newcommand{\hii}{\mbox{{\sc H ii}}}

\newcommand{\mvir}{\mbox{${M}_{\rm vir}$}}
\newcommand{\mstar}{\mbox{${M}_{\rm star}$}}
\newcommand{\rvir}{\mbox{${R}_{\rm vir}$}}
\newcommand{\ramses}{\mbox{{\sc \small Ramses}}}
\newcommand{\FRR}{$\textsf{FR}$}
\newcommand{\FRU}{$\textsf{FRU}$}

\definecolor{gray}{rgb}{0.5,0.5,0.5}

\usepackage{hyperref}
\hypersetup{colorlinks=true,citecolor=blue,linkcolor=black,filecolor=black,runcolor=black}

\begin{document}

\title{Escape fraction of ionizing photons during reionization: effects due to supernova feedback and runaway OB stars}

\author{Taysun Kimm \& Renyue Cen}
\affil{Department of Astrophysical Sciences, Princeton University, Peyton Hall, Princeton, NJ 08544, USA}

\shorttitle{Feedback-regulated escape of ionizing photons}
\shortauthors{Kimm \& Cen}

\begin{abstract}

The fraction of hydrogen ionizing photons escaping from galaxies into the intergalactic medium
is a critical ingredient in the theory of reionization. We use two zoomed-in, high-resolution (4 pc), 
cosmological radiation hydrodynamic simulations with adaptive mesh refinement to investigate 
the impact of two physical mechanisms (supernova feedback and runaway OB stars) on the escape 
fraction ($f_{\rm esc}$) at the epoch of reionization ($z\ge7$).  We implement a new, physically 
motivated supernova feedback model that can approximate the Sedov solutions at all 
(from the free expansion to snowplow) stages.
We find that there is a significant time delay of about ten million years between the 
peak of star formation and that of escape fraction, 
due to the time required for the build-up and subsequent destruction of the star-forming cloud by supernova feedback. 
Consequently, the photon number-weighted mean escape fraction for 
dwarf galaxies in halos of mass $10^8-10^{10.5}\,\msun$ is found to be $\left<\fesc\right>\sim 11\%$, 
although instantaneous values of $f_{\rm esc}>20\%$ are common  
when star formation is strongly modulated by the supernova explosions.
We find that the inclusion of runaway OB stars increases the mean escape fraction by 22\% to $\left<\fesc\right>\sim 14\%$.
As supernovae resulting from runaway OB stars tend to occur in less dense environments, 
the feedback effect is enhanced and star formation is further suppressed in halos with $\mvir\gtrsim10^9\,\msun$
in the simulation with runaway OB stars compared with the model without them.
While both our models produce enough ionizing photons to maintain a fully ionized universe at $z\le 7$ as observed, 
a still higher amount of ionizing photons at $z\ge 9$ appears necessary 
to accommodate the high observed electron optical depth inferred from cosmic microwave background observations.
\end{abstract}


\section{Introduction}

\citet{gunn65} predicted that  Ly$\alpha$ absorption would give rise to a sudden drop of 
continuum flux at wavelengths shorter than 1216 $\AA$ if a tiny amount of neutral hydrogen is present along the line of sight. 
The dramatic clearing of the Gunn-Peterson trough 
from the observation of quasars at $z\sim6$ demonstrates 
that hydrogen in the Universe is highly ionized at $z\lesssim6$ \citep{becker01,fan01,fan06}. 
Polarization signals from the comic microwave background (CMB) also suggest that 
a large fraction of hydrogen may already be ionized by $z \sim 10-12$ \citep{komatsu11,planck-collaboration13}.
Yet, the detailed processes on how reionization has occurred remain unclear.

In the standard $\Lambda$CDM universe, dwarf galaxies form early \citep[e.g.,][]{somerville03}
and could dominate the budget of hydrogen ionizing photons at the epoch of reionization. 
Photons that escape from the porous interstellar medium \citep[ISM,][]{clarke02}, 
driven by supernova (SN) explosions \citep{mckee77}, 
to the intergalactic medium (IGM) create \hii\ bubbles, which expand as more stars form.
The eventual percolation of \hii\ bubbles would mark the end of the cosmological reionization
\citep[e.g.,][]{gnedin00,mcquinn07,shin08}. 
This stellar reionization scenario has been studied extensively, both (semi-) analytically \citep[e.g.][]{madau99,miralda-escude00,barkana01,bianchi01,cen03,wyithe03,somerville03,bolton07,wyithe07,kuhlen12,robertson13} and 
numerically \citep[e.g.][]{gnedin00,razoumov02,ciardi03,fujita03,trac07,gnedin08,wise09,razoumov10,yajima11,Paardekooper13}.
It appears that dwarf galaxies are the most plausible source of the ionizing photons,
provided that the escape fraction is significant ($\fesc >10 \%$).
Active galactic nuclei also contribute to ionizing photons in both the ultraviolet (UV) and X-ray bands but
are generally believed to be sub-dominant to stellar sources 
\citep{haehnelt01,wyithe03,schirber03,faucher-giguere08a,cowie09,willott10,fontanot14}.
The strong accretion shock present in massive halos ($\mvir \gtrsim 10^{10.5}\, \msun$) 
may also produce a non-negligible amount of hydrogen ionizing photons in the vicinity of the galactic gaseous disk \citep{dopita11}.

The major uncertainty in the dwarf galaxy-driven reionization picture is the escape fraction of ionizing photons.  
Observationally, this is difficult to probe, because the hydrogen ionizing photons escaping from 
dwarf galaxies will get easily absorbed by the IGM during reionization ($z\gtrsim7$).
Besides, it requires a large sample of galaxies to obtain a statistically significant estimate of the
escape fraction ($f_{\rm esc}$). Nevertheless, it is worth noting that galaxies at higher 
redshift often exhibit a larger relative escape fraction ($f_{\rm esc}^{\rm rel}$), which is defined as the ratio of 
the escape fraction at 900$\AA$ and 1500$\AA$, than their low-$z$ counterparts \citep{siana10}. 
Observations of star-forming galaxies at $z\lesssim1$ indicate that the relative escape fraction is only 
a few percent \citep{leitherer95,deharveng01,malkan03,siana07,cowie09,bridge10,siana10}. 
The only exception reported so far is Haro 11, which shows  $f_{\rm esc}\sim 4-10\%$ \citep{bergvall06}.
On the other hand,  a non-negligible fraction ($\sim10\%$) of star-forming galaxies at $z\sim3$ reveals 
a high escape of  $f_{\rm esc}^{\rm rel} \ge  0.5$ \citep{shapley06,iwata09,nestor11,nestor13,cooke14}.
For typical Lyman break galaxies at $z\sim3$ in which 20--25\% of UV photons are escaping \citep{reddy08},
the relative fraction corresponds to a high escape fraction of $\fesc\sim0.1$.
Given that galaxies are more actively star forming at high redshift \citep[e.g.][]{bouwens12a,dunlop13},
it has been suggested that there may be a correlation between star formation rate and \fesc, 
and possibly evolving \fesc\ with redshift \citep[][]{kuhlen12}. 

Predicting the escape fraction in theory is also a very challenging task.
This is essentially because there is little understanding on the structure of the ISM at high-$z$ dwarf galaxies. 
Numerical simulations are perhaps the most suited to investigate this subject, 
but different subgrid prescriptions and/or finite resolution often lead to different conclusions. 
Using an adaptive mesh refinement (AMR) code, ART \citep{kravtsov97},  with SN-driven energy 
feedback, \citet{gnedin08} claim that the angle-averaged escape fraction increases with galaxy mass 
from $10^{-5}$ to a few percents in the range $10^{10} \lesssim M_{\rm gal} \le 4\times10^{11}$.
They attributed this trend to the fact that more massive galaxies have smaller gas-to-stellar scale-height than 
lower mass galaxies in their simulations. On the other hand, \citet{razoumov10} argue based on cosmological 
TreeSPH simulations \citep{sommer-larsen03} that more than 60\% of the 
hydrogen ionizing photons escape from dwarf galaxies in dark matter halos of $M_{\rm halo}=10^8-10^9\msun$. 
More massive halos of $10^{11}\msun$ are predicted to have a considerably smaller \fesc\ ($\lesssim 10\%$). 
A similar conclusion is reached by \citet{yajima11}.  It should be noted, however, that resolution could 
potentially be an issue in these two studies in the sense that their resolution of a few hundreds to 
thousands of parsec is unable to resolve most star-forming regions and hence capture obscuring 
column densities and a porous ISM. \citet{wise09} performed cosmological radiation hydrodynamic 
simulations employing very high resolution (0.1 pc), and found that the neutral hydrogen column 
density varies over the solid angles from $N_{\rm HI}\sim 10^{16}\, {\rm cm^{-2}}$ 
to $10^{22}\, {\rm cm^{-2}}$ with the aid of SN explosions and photo-ionization.
Because of the porous ISM, a high \fesc\ of $\sim40\%$ is achieved   
in small halos of $M_{\rm halo}=10^{7} - 10^{9.5} \msun$. 
\citet{wise14} show that an even higher fraction ($\sim 50\%$) of hydrogen 
ionizing photons escapes from minihalos of $M_{\rm halo}=10^{6.25} - 10^{7} \msun$.

Another potentially important source of ionizing radiation is runaway OB stars that are dynamically 
displaced from their birthplace. The runaway OB stars are normally defined by their peculiar motion 
\citep[$v_{\rm pec} \ge 30\, {\rm km\,s^{-1}}$,][]{blaauw61}, and roughly $30\%$ of OB stars are 
classified as runaways in the Milky Way \citep{stone91,hoogerwerf01,tetzlaff11}.
Although the fraction is still uncertain, their peculiar speed of $\left<v_{\rm pec}\right>\sim 40\,{\rm km\,s^{-1}}$ means 
that the runaway OB stars can, in principle, travel away from the birthplace by $\sim$200 pc in 5 Myrs,
making them an attractive source for the ionizing photons. 
The runaway OB stars are thought to originate from a three-body interaction with other stars in a young 
cluster \citep{leonard88}, and/or from a SN explosion of a companion in a binary system \citep{blaauw61}.
\citet{conroy12} evaluated the impact of the inclusion of runaway OB stars on \fesc\ using a simple analytic argument, 
and concluded that the runaway OB stars may enhance \fesc\ by a factor of up to $\sim4.5$ in halos with  
$M_{\rm halo}=10^8-10^9\msun$. 

The aim of this study is to investigate the importance of the aforementioned two processes 
by measuring the escape fraction from high-resolution cosmological radiation hydrodynamics simulations. 
First, given that modeling the SN explosion as thermal energy
 is well known to have the artificial radiative cooling problem \citep[e.g.][]{katz92,slyz05}, 
we expect that the role of the SN is likely to be underestimated in some cosmological simulations \citep[e.g.][]{gnedin08}. 
With a new physically based SN feedback model that captures all stages of the Sedov explosion from 
the free expansion to the snowplow phase, we study the connection between the escape of ionizing photons and feedback processes 
in dwarf galaxies. Second, we extend the idea by \citet{conroy12}, and quantify 
the impact from the runaway OB stars on reionization in a more realistic environment.

We first describe the details of our cosmological radiation hydrodynamics simulations 
including the implementation of runaway OB stars in Section~2.  
We present the feedback-regulated evolution of the escape fraction and the impact of the inclusion 
of runaway OB stars in Section~3. We summarize and discuss our findings in Section~4. 
Our new mechanical feedback from SN explosions  is detailed in Appendix.

\section{Method}

\subsection{Hydrodynamics code}
We make use of the Eulerian adaptive mesh refinement code, {\sc ramses} \citep[][ver. 3.07]{teyssier02}, to investigate 
the escape of ionizing radiation from high-$z$ galaxies. 
{\sc ramses} is based on the fully threaded oct-tree structure \citep{khokhlov98}, 
and uses the second-order Godunov scheme to solve Euler equations.
The hydrodynamic states reconstructed at the cell interface are limited using the MinMod method,
and then advanced using the Harten-Lax-van Leer contact wave Riemann solver \citep[HLLC,][]{toro94}.
We adopt a typical Courant number of 0.8. The poisson equation is solved using the adaptive particle-mesh method.
Gas can effectively cool down to $10^4$ K by atomic and metal cooling \citep{sutherland93}.
Below $10^4$ K, metal fine-structure transitions, such as {\sc [CII]} 158$\mu m$, can further lower 
the temperature down to 10 K, as in \citet{rosen95}. We set the initial metallicity to $2\times10^{-5}$, 
as primordial SNe can quickly enrich metals in mini-halos of mass $10^7\,\msun$ \citep[e.g.,][]{whalen08}, 
which our simulations cannot resolve properly.

We use the multi-group radiative transfer (RT) module developed by \citet{rosdahl13} 
to compute the photoionization by stars. 
The module solves the moment equations for three photon packets ({\sc Hii}, {\sc Heii}, and {\sc Heiii} ionizing photons) 
using a first-order Godunov method with M1 closure for the Eddington tensor. 
We adopt the Harten-Lax-van Leer \citep[HLL,][]{harten83} intercell flux function. 
Ionizing photons from each star are taken into consideration in every fine step.
Note that an advantage of  the moment-based RT is that it is not limited by the number of sources.
The production rate of the ionizing photon varies with time for a given initial mass function \citep[IMF,][see also \citealt{rosdahl13}]{leitherer99}. The majority of the ionizing photons are released in $\sim$ 5 Myr of stellar age.
We adopt the production rate equivalent to that of Kroupa IMF \citep{kroupa01} 
from the {\sc Starburst99} library \citep{leitherer99}\footnote{Note that we use the Chabrier IMF to 
estimate the frequency of SN explosions. We choose the number of ionizing  photons equivalent to that of the Kroupa IMF, 
because the models with the Chabrier IMF is not yet available in the {\sc Starburst99} \citep{leitherer99}}.
The radiation is coupled with gas via photo-ionization and photo-heating,
and a set of non-equilibrium chemistry equations for {\sc Hii}, {\sc Heii}, and {\sc Heiii} 
are solved similarly as in \citet{anninos97}. We assume that photons emitted by recombination are 
immediately absorbed by nearby atoms (case B).
The speed of light is reduced for the speed-up of the simulations by 0.01 \citep[e.g.][]{gnedin01}.
This is justifiable because we are mainly interested in {\it the flux} of escaping photons at the virial sphere.

\begin{table}
   \caption{Summary of cosmological simulations}
      \label{table1}
      \centering
   \begin{tabular}{@{}cccccccc} 
   \hline
   \hline
Model &  SNII & RT & Run-       & $\Delta x_{\rm min}$ & ${m_{\rm star,min}}$ &  $m_{\rm dm}$  \\
            &           &       & aways     &  [pc]                              &    [$\msun$]                & [$10^5\,\msun$]  \\
\hline
FR & $\checkmark$ &  $\checkmark$ & -- &  4.2 & 49 &  1.6 \\
FRU &$\checkmark$ & $\checkmark$ & $\checkmark$ & 4.2 & 49 & 1.6  \\
\hline
   \end{tabular}
\end{table}

\begin{figure}
   \centering
   \includegraphics[width=7.5cm]{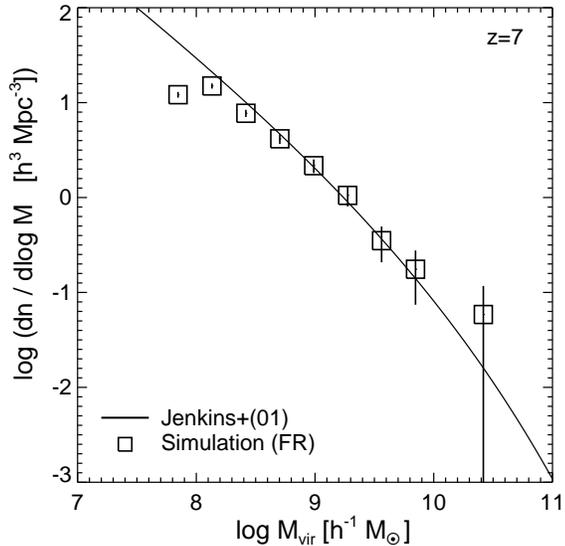}
   \caption{Dark matter halo mass function from the zoomed-in region of the \FRR\ run at $z=7$.
   Comparison with \citet{jenkins01} mass function at the same epoch indicates that our simulated volume 
   represents the average region of the universe. }
   \label{fig:mf}
\end{figure}

\subsection{Cosmological Simulations}

We carry out cosmological simulations to investigate 
the escape fraction in realistic environments. For this purpose, we generate the initial condition 
using the {\sc music} software \citep{hahn11}, with the WMAP7 cosmological parameters \citep{komatsu11}:
$(\Omega_{\rm m}, \Omega_{\Lambda}, \Omega_{\rm b}, h, \sigma_8, n_s  = 0.272, 0.728, 0.045, 0.702, 0.82, 0.96)$.
A large volume of $(25\,{\rm Mpc} \, h^{-1})^3$ is employed to include the effect of the large-scale tidal field. 
To achieve high mass resolution, we first run dark matter-only simulations with 256$^3$ particles, 
and identify a rectangular region of $3.8\times4.8\times9.6$ Mpc (comoving)
that encloses two dark matter halos of $\simeq 1.5\times 10^{11} \msun$ at $z=3$.
Then, we further refine the mass distribution of the zoomed-in region, such that the mass of a dark matter particle 
is $m_{\rm dm}=1.6\times10^5\,\msun$, which corresponds to 2048$^3$ particles in effect.
Despite that we purposely select the region in which two massive dark matter halos are present at $z=3$,
a comparison with the number of dark matter halos per volume predicted by \citet{jenkins01} shows that 
our simulated box represents an average region of the universe at $z=7$  (Figure~\ref{fig:mf}). 

\begin{figure}
   \centering
      \includegraphics[width=8.6cm]{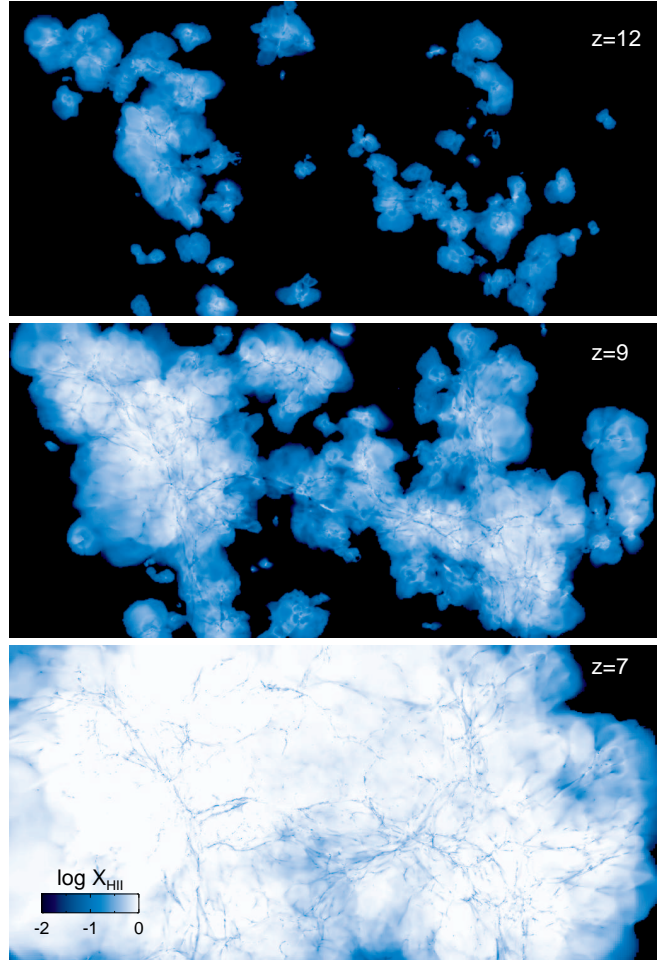}
      \caption{Expansion of the \hii\ bubble in a cosmological simulation (\FRR). Three panels show the evolution of 
      the density-weighted fraction of ionized hydrogen of the zoomed-in region. The horizontal size of the figure is 
      9.5 Mpc (comoving).}
            \label{fig:hii}
\end{figure}

The level of the root grid in the zoomed-in region is 11, consistent with the dark matter resolution.
Further 12 levels of refinement are triggered if the dark matter plus baryon mass in a cell exceeds 
8 times the mass of a dark matter particle. We keep the minimum physical size of a cell to 
 $\Delta x_{\rm min}=25\,{\rm Mpc} \, h^{-1}/ 2^{23} = 4.2\,{\rm pc}$ over the entire redshift. However, this refinement 
criterion is not optimized to resolve the structure of the ISM, unless extremely high mass resolution is adopted. 
For example, for a gas cell of $n_{\rm H}=10\, {\rm cm^{-3}}$, the criterion will come into play only if the size of 
the cell is larger than $\sim$ 160 pc.  In order to better resolve the structure of the ISM, 
we enforce a cell with $n_{\rm H}\ge 1\, {\rm cm^{-3}}$ to be resolved on $8 \Delta x_{\rm min}=34\,{\rm pc}$.
In a similar context, we apply more aggressive refinement criterion for the star-forming gas 
in such a way that gas with $n_{\rm H}=100\, {\rm cm^{-3}}$ ($800\,{\rm cm^{-3}}$) is always 
resolved on a 8.5 pc (4.2 pc) cell.
We adopt very high stellar mass resolution of $\approx 49\,\msun$.
This means that a star particle with the minimum mass will produce a single SN event for the Chabrier IMF.

We run two sets of cosmological simulations, \FRR\ and \FRU, with the identical initial condition down to $z=7$.
Both runs include star formation, metallicity-dependent radiative cooling \citep{sutherland93,rosen95}, 
thermal stellar winds, mechanical feedback from SN explosions, and photoionization by stellar radiation.
The runaway OB stars are included only in the \FRU\ run. In Figure~\ref{fig:hii}, we show an example of the 
growth of \hii\ bubbles in the \FRR\ run. Our simulated region is nearly ionized at $z=7$.

Dark matter (sub) halos are identified using the {\sc Amiga} halo finder \citep[{\sc Ahf},][]{gill04,knollmann09}.
{\sc Ahf} first constructs the adaptive meshes based on the particle distribution, finds the density minima,
and determines physical quantities based on a virial overdensity ($\Delta_{\rm vir}$).
Gravitationally unbound particles are removed iteratively if they move faster than the local escape velocity 
during this procedure. The virial radius is defined such that 
the mass enclosed within the virial sphere is the virial overdensity times the critical density of the universe times the volume, 
i.e. $\mvir(z) = \Delta_{\rm vir}(z) \rho_{\rm crit}(z) 4 \pi r_{\rm vir}^3 / 3$.
We take $\Delta_{\rm vir}=177$, appropriate for a $\Lambda$-dominated universe at $z>6$ \citep{bryan98}.
This results in 796, 443, and 183 dark matter halos of mass $\mvir\ge10^{8}\,\msun$ immune to the contamination 
by coarse dark matter particles ($m_{\rm dm} > 1.6\times10^{5}\,\msun$) at $z=7$, 9, and 11, respectively.

\subsection{Star Formation and Feedback}

Stars form in a very dense, compact molecular core. 
Infrared extinction maps of nearby interstellar cores indicate that their size ranges from 0.01 to 0.4 pc
\citep[e.g.][]{alves07,konyves10}, which is difficult to resolve in current cosmological simulations.
Nevertheless, studies of gravitational collapse in converging flows \citep{gong11} seem to suggest that 
a gravitationally bound cloud is likely to experience runaway collapse no matter how the collapse is initiated.  
In a similar spirit, we assume that stars would form in a cell if the following conditions are met simultaneously
\citep[e.g.][]{cen92}:
\begin{itemize}
\itemsep0em
\item[1.] the flow is convergent ($\vec{\nabla}\cdot \rho {\vec  v} <0$) ,
\item[2.] the cooling time is shorter than the dynamical time, 
\item[3.] the gas is Jeans unstable,  and
\item[4.] the number density of hydrogen exceeds the threshold density $n_{\rm th}={\rm 100 \,cm^{-3}}$.
\end{itemize}
The last condition is motivated by the density of  a Larson-Penston profile \citep{larson69,penston69} at $0.5\Delta x$,
 $\rho_{\rm LP}\approx8.86 c_s^2 / \pi\,G\,\Delta x^2$, where $c_s$ is the sound speed and $\Delta x$ is 
 the size of the most refined cell. 
 Star particles are created based on the Schmidt law \citep[][]{schmidt59}, 
 $ \dot{\rho}_{\star} = \epsilon_{\rm ff} \, \rho_{\rm gas} \, / \, t_{\rm ff} $,  assuming that 2\% of the star-forming 
 gas ($\epsilon_{\rm ff}$)  is converted into stars per its free-fall time ($t_{\rm ff}$) \citep{krumholz07,kennicutt98}.  
 The mass of each star particle is determined as $m_\star=\alpha\, N_p \rho_{\rm th} \, \Delta x_{\rm min}^3 $, 
 where $\rho_{\rm th}$ is the threshold density for star formation, 
$\Delta x_{\rm min}$ is the size of the most refined cell,  and $\alpha$ is a parameter that 
controls the minimum mass of a star particle. $N_p$ is the number of star particles to be formed in a cell,
which is drawn from a Poisson random distribution, $P(N_p) = (\lambda ^{N_p} / N_p! ) \exp\left(-\lambda\right)$.
Here the Poissonian mean ($\lambda$) is computed as 
$\lambda \equiv \epsilon_{\rm ff} \left({\rho\Delta x^3}/{m_{\rm \star,min}}\right) \left( {\Delta t_{\rm sim}}/{t_{\rm ff}}\right), $
where $\Delta t_{\rm sim}$ is the simulation time step, and $m_{\rm \star,min}$ is the minimum stellar mass (i.e. $N_p=1$).

We describe the SN feedback using a new physical model which captures 
the SN explosion at all stages from the early free expansion to the final momentum-conserving snowplow phase.
Briefly, we deposit radial momentum to the cells affected by supernova feedback, conserving energy appropriately.
The amount of input momentum is determined by the stage the blast wave is in, which in turn is dependent upon the 
physical condition  (density and metallicity) of the gas being swept up and simulation resolution. 
The virtue of our scheme is that an approximately (within 20\%) correct  amount of momentum is imparted to the 
surrounding gas regardless of the resolution. Thus, this prescription should be useful to cosmological simulations,
especially those with finite resolution that potentially suffer from the artificial radiative cooling. 
The details of our implementation and a simple test are included in the Appendix.

The frequency of a SN per solar mass is estimated assuming the Chabrier IMF \citep{chabrier03}.
For the simple stellar population with a low- (high-) mass cut-off of 0.1 (100) \msun, 
the total mass fraction between 8 to 100 \msun\ is 0.317, and the mean SN progenitor mass is 15.2 \msun\ on the zero-age main sequence.
At the time of the explosion, we also deposit newly processed metals into the surrounding. 
The mass fraction of newly synthesized metals in stellar ejecta is taken to be 0.05 following \citet{arnett96}.
A star particle is assumed to undergo the SN phase after the main sequence lifetime of the mean SN progenitor \citep[10 Myr,][]{schaller92}. As discussed in \citet{slyz05}, allowing for the delay between the star formation and explosion 
(i.e. stellar lifetimes) is crucial to the formation of hot bubble in the ISM. 
We find that the physically based SN feedback employed in this study drives stronger galactic winds 
than the runs with thermal feedback or kinetic feedback that are valid only under certain conditions
 \citep[][see below]{dubois08}. 
Stellar winds from massive stars are modeled as thermal input, based on \citet{leitherer99}.

\subsection{Runaway OB Stars}

Our implementation of runaway OB stars is largely motivated by \citet{tetzlaff11},
who compiled candidates of runaway stars younger than 50 Myr for the 7663  {\it Hipparcos} sample. 
By correcting the solar motion and Galactic rotation, they found that the peculiar space velocity of the stars 
may be decomposed into two Maxwellian distributions intersecting at 28 ${\rm km\,s^{-1}}$.
Assuming that each Maxwellian distribution represents a kinematically distinctive population, 
they estimated the fraction of the runaways to be $\sim 27.7\%\pm 1.9$ for the sample 
with full kinematic information. The dispersion of the Maxwellian distribution is 
measured as 24.4 ${\rm km\, s^{-1}}$ for the high-velocity group.

Since either runaway OB stars formed through the explosion of a SN in a binary or 
those dynamically ejected in a cluster are not resolved in our simulations, 
we crudely approximate this by splitting a star particle into a normal (70\% in mass) 
and a runaway particle (30 \%) at the time of star formation. While the initial velocity of the normal star is chosen 
as the velocity of the birth cloud, we add a velocity drawn from the Maxwellian distribution 
on top of the motion of the birth cloud for runaway particles. To do so, we generate the distribution following 
the Maxwellian with the dispersion of $\sigma_v = 24.4\,{\rm km\,s^{-1}}$ and the minimum space 
velocity of $v_{\rm 3D}=28 \,{\rm km\,s^{-1}}$ using the rejection method \citep{press92}. The direction of the 
runaway motion is chosen randomly for simplicity. A similar approach is taken by \citet{ceverino09} to 
study the formation of disk galaxies in a cosmological context.

\subsection{Estimation of Escape Fraction}

The fraction of escaping ionizing photons ($f_{\rm esc}$) is measured by comparing the photon flux at the virial radius 
and the photon production rate from young massive stars.
Since the speed of light is finite, there is a small delay in time between the photons produced by the stars and 
the photons escaping at the virial sphere. In order to take this into account, we use the photon production rate at earlier time ($t-r_{\rm vir}/c'$), 
where $c'$ is the reduced speed of light used in the simulations. The escape fraction is then computed as
\begin{equation}
f_{\rm esc}(t) \equiv \frac{\int d\Omega\, \vec{F}_{\rm ion}(t) \cdot \hat{r} ~\Theta(\vec{F}_{\rm ion}\cdot \hat{r})}{\int dm_* \, \dot{N}_{\rm ion} (t-r_{\rm vir}/c')},
\label{eq:fesc}
\end{equation}
where $\vec{F}_{\rm ion}$ is the ionizing photon flux, $d\Omega$ is the solid angle, $m_*$ is the mass of each star particle,
$\dot{N}_{\rm ion}(t)$ is the photon production rate of a simple stellar population of age $t$ per solar mass, 
and $\Theta$ is the Heaviside step function. Here, we approximate the delay time  to be a constant, $r_{\rm vir}/c'$, 
for each halo assuming that the central source is point-like. 
Since only outflowing photons are considered in Equation~\ref{eq:fesc},
we find that  a minor fraction ($\sim 5\%$) of galaxies exhibit $f_{\rm esc}$ greater than 1.
This happens mostly when there is little absorbers left in the halo after disruptive SN explosions.
In this case, we randomly assign $f_{\rm esc}$ between 0.9 and 1.0.
We confirm that the photon production rate-averaged escape fraction, which is the most important quantity in this study, 
is little affected by this choice even if the net flux is used, and thus we decide to take a simpler method. 

Dust can also affect the determination of the escape of the hydrogen ionizing photons. 
However, given that our simulated galaxies are very 
metal-poor ($0.002-0.05\,Z_{\odot}$) and galaxies with lower metallicity have a progressively lower amount of 
dust \citep{lisenfeld98,engelbracht08,galametz11,fisher13},  it is unlikely that dust decreases the escape 
fraction substantially. Thus, we neglect the absorption of hydrogen ionizing photons by dust in this study.

\section{Results}
\subsection{Feedback-regulated Escape of Ionizing Photons}

\begin{figure}
   \centering
            \includegraphics[width=8.5cm]{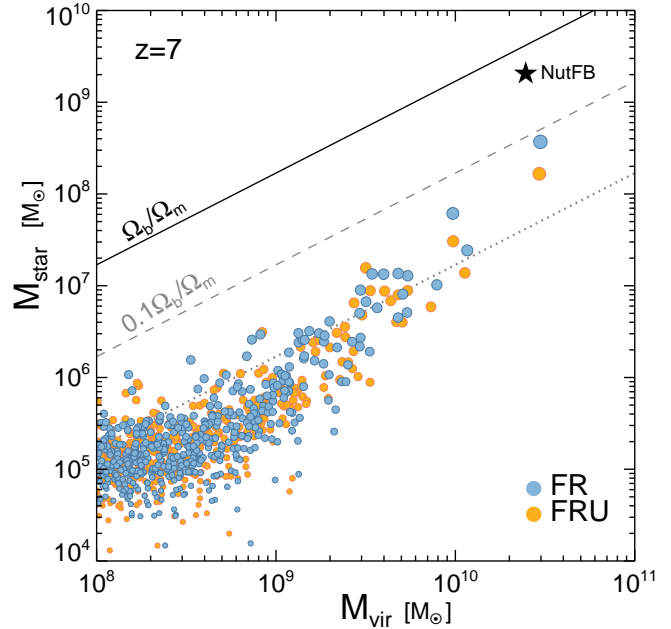}
   \caption{The baryon-to-star conversion efficiency at $z=7$ from the \FRR\ (blue) and the \FRU\ (orange) runs. 
   Only central galaxies are shown. The cosmic mean    ($\Omega_{\rm b}/\Omega_{\rm m}=0.165$) is 
   shown as a black solid line. Also included as a star is the stellar fraction measured from 
   the NutFB simulation \citep{kimm11b}.
   Our mechanical feedback from SN explosions is more effective 
   at regulating star formation, compared with previous studies injecting thermal or kinetic energy (see the text).
   }
   \label{fig:mstar}
\end{figure}

\begin{figure*}
   \centering
      \includegraphics[width=17cm]{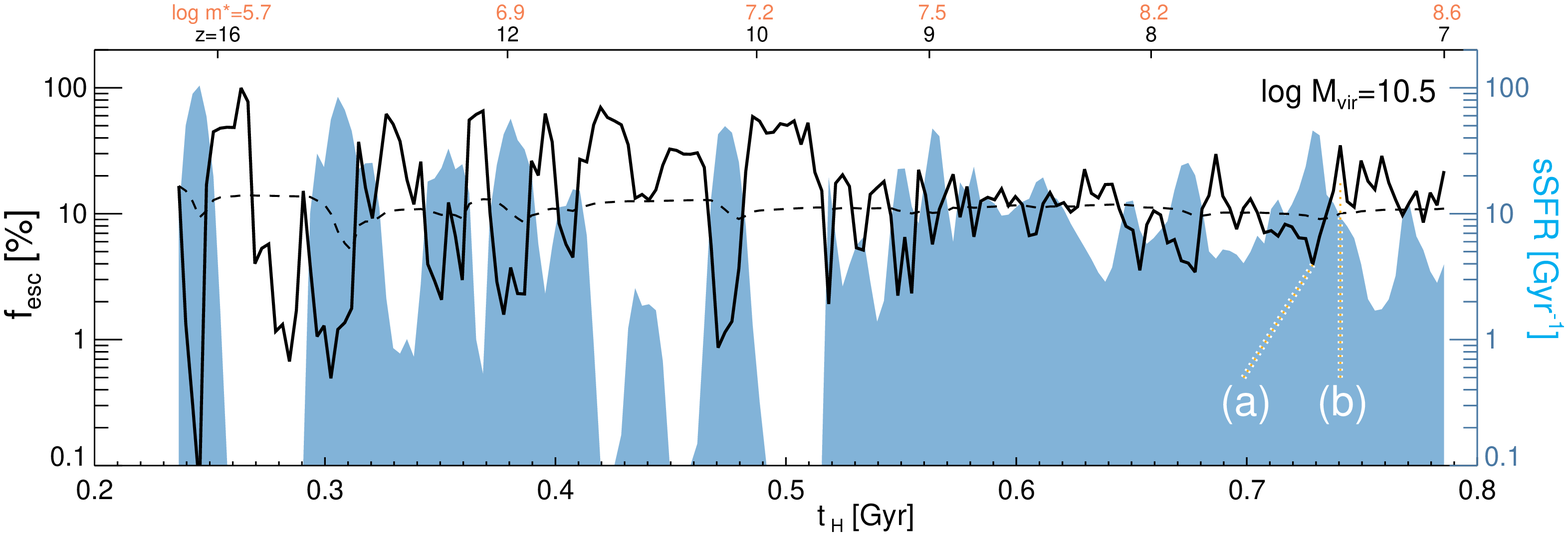}
      \includegraphics[width=18cm]{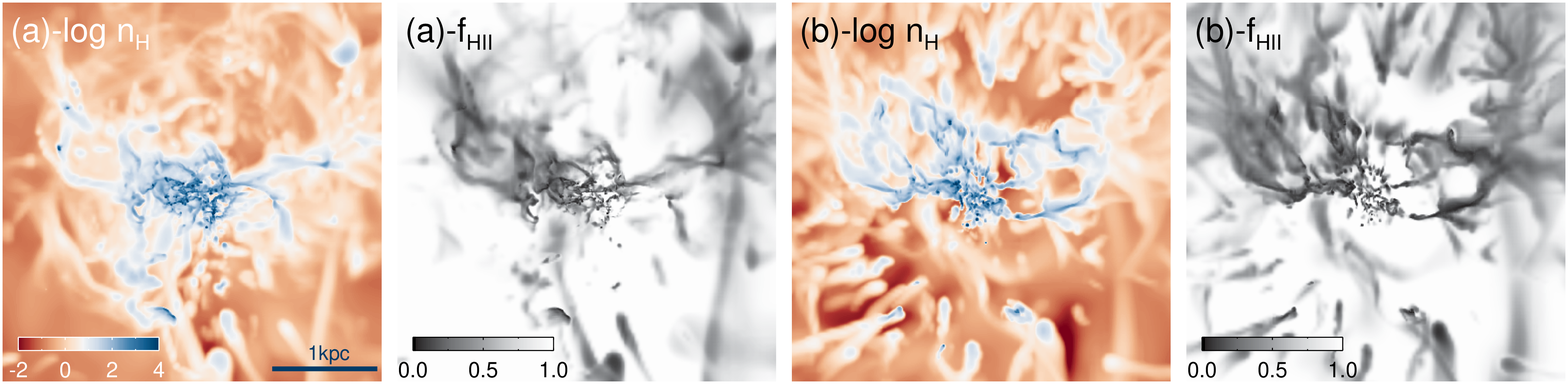}
      \includegraphics[width=17cm]{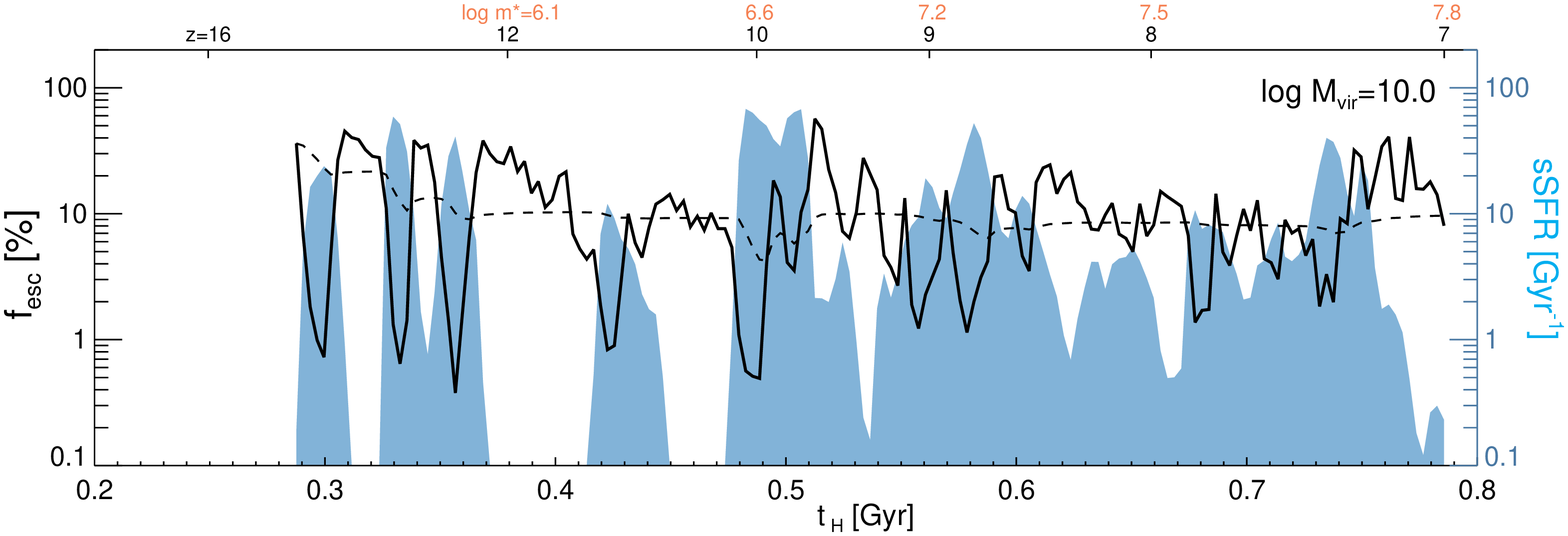}
      \caption{Evolution of the escape fraction (\fesc) and specific star formation rate (sSFR) in two massive 
      halos from the \FRR\ run. Black solid lines in the top and bottom panels indicate the escape fraction measured 
      at the virial radius at each snapshot as a function of the age of the universe. We denote the logarithmic stellar mass 
      at different times by orange text.
      Black dashed lines correspond to the photon number-weighted average of \fesc\ by that time (\fescg). 
      Blue shaded regions display the sSFR in ${\rm Gyr^{-1}}$. One can see that there is a delay between 
      the peak in \fesc\ and sSFR due to the 
      delay in the onset of the strong outflow. The middle panels show 
      an example of this delay identified in the top panel (a,b). The projected density of gas and the fraction of 
      ionized hydrogen are shown in both cases, as indicated in each panel. Interestingly, the volume filling 
      fraction of the neutral hydrogen within 0.2 \rvir\ is found to be 25\% large in the snapshot (b), indicating that 
      \fesc\ depends not only by the volume-filling, circumgalactic neutral gas, but also dense star
      forming gas.  We do not display the physical quantities if $M_{\rm vir}\le10^8\,\msun$.
      }
            \label{fig:ex}
\end{figure*}

\begin{figure*}
   \centering
            \includegraphics[width=8.2cm]{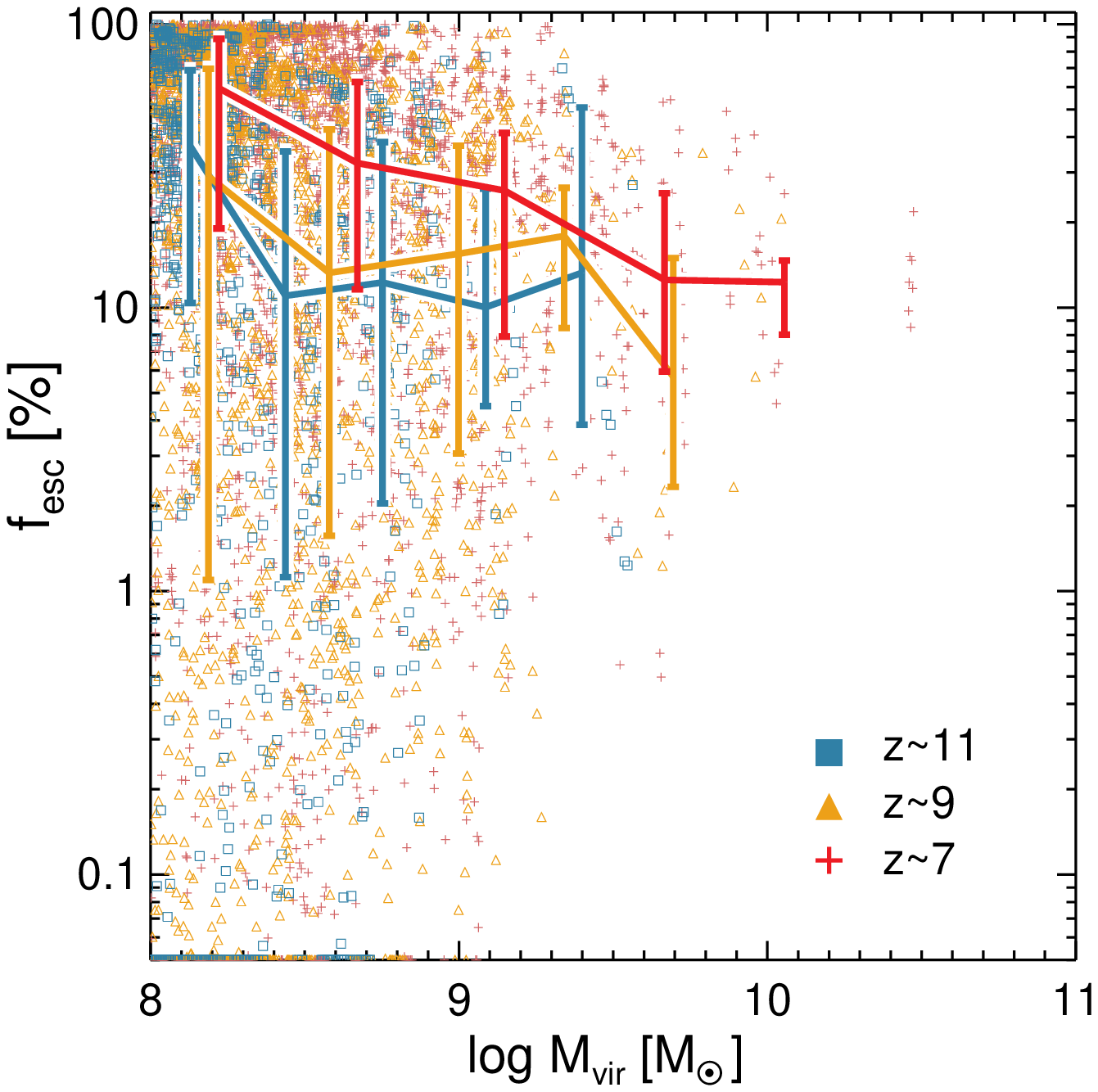}
                  \includegraphics[width=8.2cm]{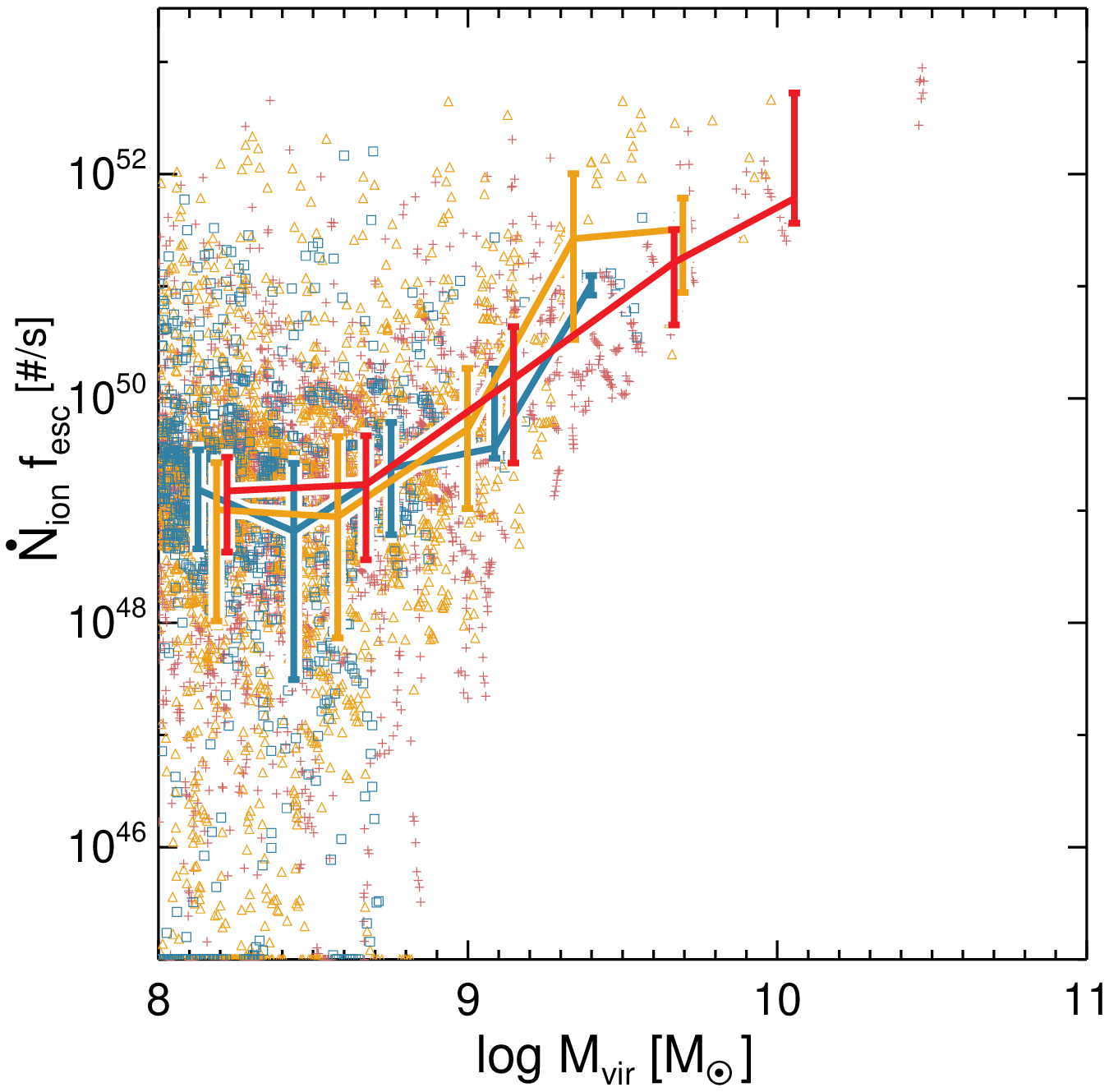}
   \caption{ 
   {\it Left}: Escape fraction measured at the virial radius at three different redshifts from the \FRR\ run.
   Different redshifts are shown as different colors and symbols, as indicated in the legend. 
   To increase the statistical significance, we combine the results 
   from seven consecutive snapshots for each redshift. Solid lines indicate the median, and error bars show 
   the interquartile range. Although there is a large scatter,  more than 50\% of the galaxies reveal $\fesc \gtrsim 10\%$.
   {\it Right:} Photons escaping per second through the virial sphere.  
   }
   \label{fig:fesc_stat}
\end{figure*}

Cosmological hydrodynamics simulations often suffer from the artificial over-cooling problem 
in forming disk galaxies \citep[e.g.][]{kimm11b,hummels12}, mainly because the energy from SN 
explosions is radiated away before it is properly transferred to momentum due to inadequate resolution
of the multi-phase ISM.  This directly affects the escape of ionizing photons. Motivated by this challenge, 
we have implemented a SN feedback scheme that reasonably approximates the Sedov blast waves 
from the free expansion to snowplow stages. In Figure~\ref{fig:mstar}, we present the baryon-to-star 
conversion efficiency ($f_{\star}\equiv M_{\rm star}/(\Omega_{\rm b} M_{\rm vir}/\Omega_{\rm m}$)
of the central galaxies in dark matter halos at $z=7$ from the \FRR\ run.  
It shows that our new physically motivated SN feedback is very effective at suppressing star formation. 
For example, the most massive halo with $M_{\rm vir}\sim 3\times10^{10}\,\msun$ at $z=7$ shows $f_{\star}\approx0.08$. 
Although the direct comparison may be difficult due to a different initial condition used, 
it is worth noting that the conversion efficiency is about a factor of 7 smaller than 
that found in the {\sc NutFB} run \citep[][see Fig.13]{kimm11b}, shown as a star in Figure~\ref{fig:mstar}.  
We note that the momentum input from SN explosions used in the {\sc NutFB} run is a factor of $3-4$ smaller 
compared with that at the end of the cooling phase \citep[see Appendix,][]{blondin98}.
For lower mass halos, the conversion efficiency is found to be even lower, 
reaching $M_{\rm star} / M_{\rm vir} \lesssim 0.01 \, \Omega_{\rm b} / \Omega_{\rm m} $ at $M_{\rm vir} \sim 10^9\,\msun$.
It is also interesting to note that the conversion 
efficiency at $M_{\rm vir}\ge 10^{10}\msun$ also agrees reasonably well within error bars with the 
semi-analytic results obtained to reproduce the observed stellar mass function, star formation rate, and 
cosmic star formation rate density  \citep[e.g.,][Figure~7]{behroozi13}.
As the feedback becomes more effective and fewer stars are formed, the stellar metallicity of these high-$z$ galaxies 
would be lower. 
We find that the most massive galaxy in our $z=7$ sample ($M_{\rm star}=4\times10^8\,\msun$) 
has a stellar metallicity of 0.05 $Z_{\rm \odot}$. 
This is at least factor of 2--3 smaller than the prediction by \citet{finlator11} at the same epoch.  
\citet{kimm13} also investigated UV properties of $z=7$ galaxies of stellar mass
$5\times10^8 - 3\times10^{10}\,\msun$ using a SN energy-based feedback scheme,
and found that stellar metallicities are generally higher than those found in the \FRR\ run. 
\citet{kimm13} found that the stellar metallicity for galaxies of mass $4\times 10^{8}\msun$
falls in the range of $0.1-0.5Z_{\rm \odot}$. 
The gas metallicities ($Z_{\rm gas}$) are also different in the two simulations.
The gas metallicity of the ISM within $2.56$ kpc for the $4\times 10^{8}\msun$ galaxies
is $0.083Z_{\rm \odot}$ in the FR run, which is about a factor of 3 lower, on average, 
than that of \citet{kimm13}  ($Z_{\rm gas}=0.1-0.7Z_{\rm \odot}$).
These comparisons lead us to conclude that our physically based feedback scheme is effective in 
alleviating the overcooling problem.

One may wonder whether stars form inefficiently in these small haloes 
($10^8\lesssim M_{\rm vir} \lesssim 10^9\,\msun$) 
because gas accretion is suppressed due to the ionizing background radiation \citep{shapiro94,thoul96,gnedin00b,dijkstra04,sobacchi13,noh14}. 
However, this is unlikely the case, given that galaxies in the atomic cooling halos 
are fed mainly by dense filaments and satellites at high redshift \citep[e.g.,][]{powell11}, 
which are self-shielded from the background radiation \citep{faucher-giguere10,rosdahl12}.
Even in the absence of the self-shielding, \citet{geen13} find no clear sign that reionization suppresses
star formation in such halos at $z>6$. \citet{wise14} also show that the fraction of baryons 
in a $10^8$-$10^9\,\msun$ halo is reduced only by less than a factor of two compared with the cosmic mean in 
their cosmological radiation hydrodynamics simulations with thermal supernova feedback and reionization.
Indeed, we confirm that our mechanical supernova feedback is 
primarily responsible for the low conversion efficiency by directly comparing the stellar mass of the dwarf galaxies 
between the simulations with and without ionizing radiation (see the Appendix).

We now present the time evolution of star formation rate and ionizing photon escape fraction
of two randomly chosen relatively massive galaxies in Figure~\ref{fig:ex}.
The plot corroborates that the feedback from stars governs the evolution of galaxies. 
The top and bottom panels show the evolution of specific star formation rate 
(sSFR$ \equiv \dot{M}_{\rm star}/M_{\rm star}$) and instantaneous \fesc\ of the central galaxy  
in dark matter halos of mass $3\times10^{10}$ and $10^{10}\,\msun$, respectively.  
The SFR is computed by averaging the mass of newly formed stars over 3 Myr.
It is evident that star formation is episodic on a time scale of $10-30$ Myr with both
the frequency and oscillation amplitude decreasing with increasing stellar mass. 
This means that SN explosions effectively 
control the growth and disruption of star-forming clouds.
When the galaxies are small ($t_{\rm H} \lesssim 0.5\, {\rm Gyr}$), the explosions even completely 
shut down the star formation across the galaxies, as stars form only in a few dense clouds. 
During these quiet periods, \fesc\ is kept high ($\fesc \gtrsim 0.2$). On the other hand, massive 
galaxies contain  many star-forming clumps,  as can be seen in the projected density plot (middle row).
The fact that the episodic star formation history becomes more smooth at late times indicates that 
these clumps are not entirely susceptible, but somewhat resilient to the SN explosions 
arising from neighboring star clusters.

More importantly, we find that there is a time delay between the peak of \fesc\ and sSFR. This is 
essentially because massive stars with $M\approx15\,\msun$ explode $\sim$10 Myr after their 
birth in our simulation. Let us suppose a dense cloud that just begins to form stars. 
Since the gas flow is usually convergent in these regions, the density of the gas will rise with time, and 
so does the SFR. This means that more and more massive stars will explode as time goes on.
Once enough SNe that can significantly redistribute the birth cloud go off, 
SFR will begin to drop, and \fesc\ will increase. Note that the increase in the number 
of SNe continues even after the peak of SFR, as massive stars live $\sim$10 Myr.
Once the massive stars formed at the peak of SFR evolve off, star formation 
will be further suppressed as a result of the destruction of the star-forming clouds, and strong 
outflows are likely to be produced, thus maximizing \fesc.
Therefore, the time delay stems from the interplay between the build-up of a non-coeval star cluster  
and subsequent SN explosions after the lifetime of the massive stars ($\sim$ 10 Myr).
The projected density distributions of gas at two snapshots, 
one of which displays the peak in sSFR (a) and the other shows the peak in \fesc\ (b), 
substantiates that it is indeed the strong outflow that elevates \fesc\ (middle row).
When sSFR is at the peak value, the central galaxy appears relatively quiet (panel-(a)), whereas  
strong outflows are seen when \fesc\ is highest and sSFR drops rapidly (panel-(b)).
As one can read from the figure, this mis-match of SFR and \fesc\ means that 
a large amount of ionizing photons at the peak of SF are absorbed by their birth clouds.
Although \fesc\ is high in the early time ($t_{\rm H} \lesssim 0.5\,{\rm Gyr}$), the photon number-weighted 
mean \fesc\ (dashed lines) stays at around $10\%$ level in these two examples.

\begin{figure*}
   \centering
            \includegraphics[width=8.6cm]{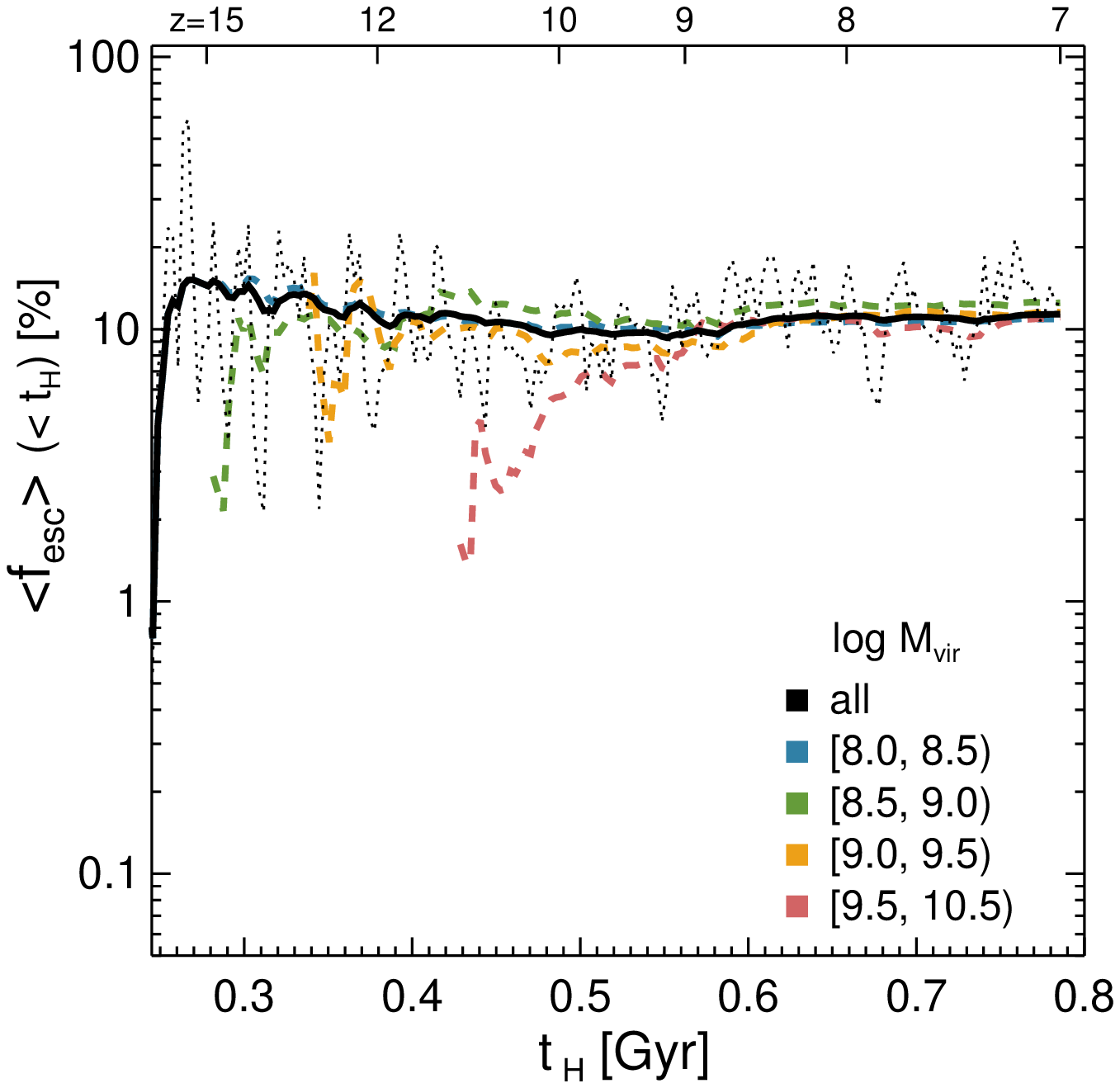}
                        \includegraphics[width=8.6cm]{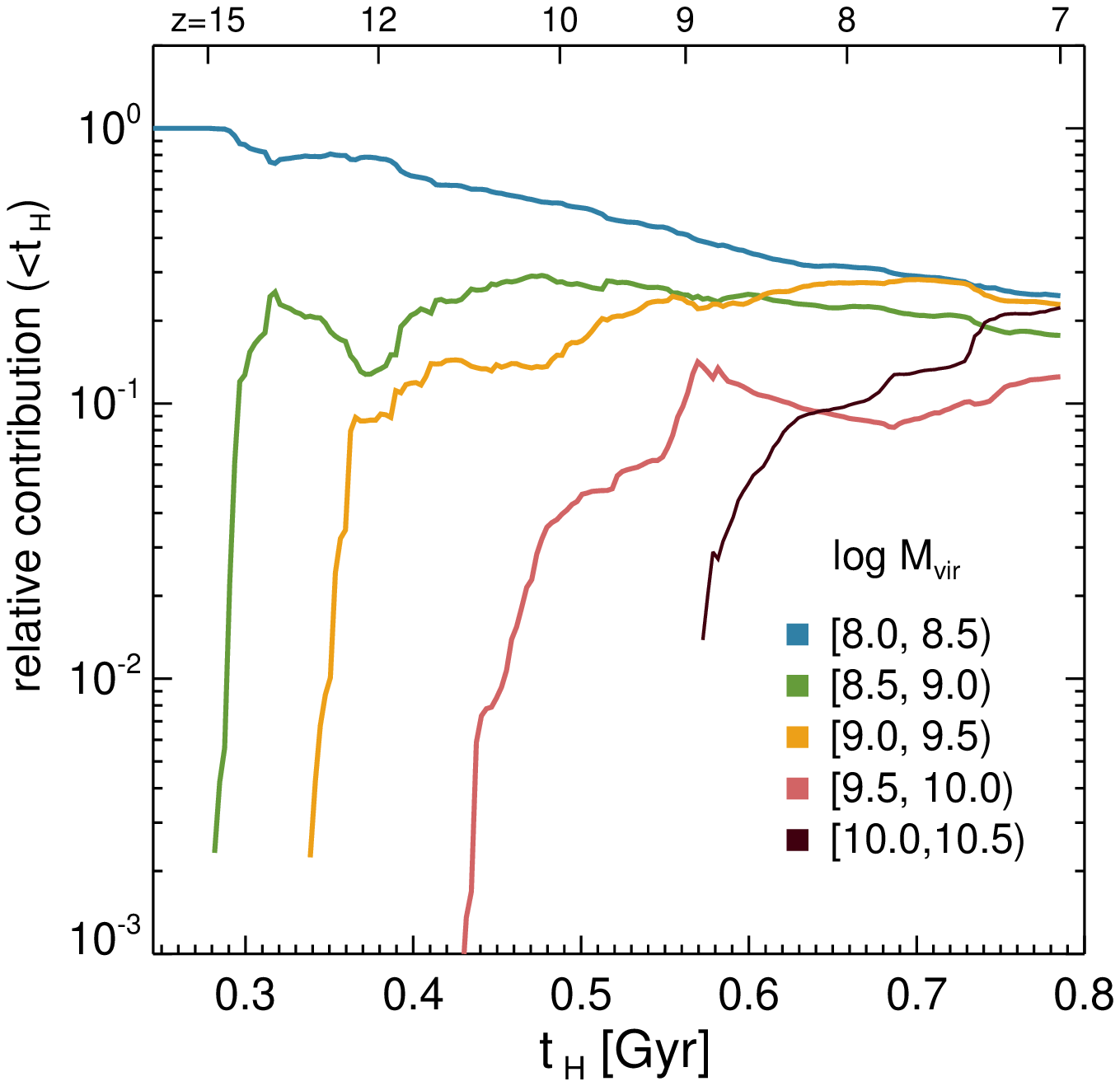}
   \caption{
   {\it Left:} Photon production rate-weighted escape fraction, $\fescg$, 
   averaged over the age of the universe ($t_{\rm H}$) in the \FRR\ run. 
   The effective escape fraction in different halo mass bins is shown as different color codings, as indicated in the legend.
   We also display the photon rate-averaged escape fraction of the whole sample at each 
   snapshot ($\fescg(t)$) (black dotted line), as opposed to the time-averaged quantities (solid and dashed lines). 
  We find the effective escape fraction to be $\sim$10\%, regardless of the halo mass and redshift. 
   Altogether, 11.4\% of the photons produced until $z=7$ have escaped from halos of $\mvir\ge10^8\,\msun$.
   {\it Right:} Relative contribution of halos of different mass ranges to the 
   total number of ionizing photons measured at the virial radius. The contribution is computed by taking into 
   account the cumulative number of photons produced and the cumulative number of photons escaped from 
   halos of relevant mass range until $t\le t_{\rm H}$.  
   }
   \label{fig:fesc_wei}
\end{figure*}

We present  statistical results of the escape fraction in Figure~\ref{fig:fesc_stat}.
Since there are a limited number of galaxies in our simulated volume and \fesc\ varies 
significantly on $\sim$10 Myrs, we compute the median and interquartile range of \fesc\ by combining 
the results from seven consecutive snapshots spanning 21 Myrs.  Several features can be gleaned from this figure.
First, although there is a considerable scatter, high-$z$ galaxies exhibit a high \fesc\  on the order of 10\%,
which is normally required by semi-analytic calculations of reionization to ionize the universe 
by $z\sim6$ \citep{wyithe07,shull12,robertson13}.
Second, there is a hint that photons can escape more easily in the galaxies hosted by lower mass halos.
We attribute this to the fact that feedback from stars efficiently destroys a few star-forming clouds that are 
responsible for the total SF in smaller halos, as opposed to larger ones in which young massive stars are 
buried in many star-forming clouds that are relatively resilient to the SN feedback arising 
from neighboring star clusters.
As shown in the top and bottom panels of Figure~\ref{fig:ex},
when galaxies are small, the entire star formation can be suppressed due to the energetic outflows driven by 
SN explosions.
Third, we find that \fesc\ is slightly higher at lower redshift for a given halo mass, consistent with \citet{Paardekooper13}.
This is essentially because the mean density of the gas is smaller at lower redshift, and the impact from SNe becomes 
more effective.

Note that high \fesc\ does not necessarily mean that more photons would leave their host halo. 
Star clusters older than $\sim$ 5 Myr would not contribute 
significantly to the total ionizing photon budget even if their \fesc\ is 1. The more relevant quantity for  
reionization should take into account the photon production rate, and we find that the (weak) redshift 
dependence of \fesc\ disappears when the photon escape rate is plotted (right panel in Figure~\ref{fig:fesc_stat}).
Since the instantaneous measurement of \fesc\ could be misleading,
we also present the photon production rate-weighted, time-averaged escape fraction, 
$\fescg (\le t_{\rm H}) \equiv  \int_0^{t_{\rm H}} \dot{N}_{\rm ion}(t) f_{\rm esc}(t) dt /  \int_0^{t_{\rm H}}  \dot{N}_{\rm ion}(t) dt,$
in Figure~\ref{fig:fesc_wei} (left panel). 
This is a better quantity to be used for the semi-analytic calculations 
of reionization than \fesc\ from Figure~\ref{fig:fesc_stat}.
Overall, we find that the time-averaged escape fraction at $z=7$ is around $\sim$ 10\%, 
regardless of the halo mass in the range considered.
Also included as the black dotted line in Figure~\ref{fig:fesc_wei} is the photon production rate-weighted average of \fesc\ 
of all the samples at different times ($\fescg(t)$). Again, the value is found to fluctuate around 10\%, 
but no clear sign of redshift dependence is detected.   

The relative contributions from halos of different masses to the total escaping ionizing photons are 
compared in Figure~\ref{fig:fesc_wei} (right panel).
As the small structures form first in the $\Lambda$CDM universe, the small halos of mass 
$\mvir \le 10^{8.5}\,\msun$ dominate down to $z\sim9$. 
More massive halos and galaxies emerge later, and their cumulative contribution 
becomes comparable with that of the smallest halos ($\mvir \le 10^{8.5}\,\msun$) by $z=7$. 
In our simulations, 14 most massive halos supply more ionizing photons than 556 smallest halos with $\mvir \le 10^{8.5}\,\msun$ at $z=7$.
This is mainly because $f_{\star}$ is much higher in the more massive halos than 
in the small halos, while the effective escape fraction is similar.
The typical number of escaping photons per second in halos with $\mvir\sim10^{8.5}\,\msun$ is 
$f_{\rm esc}\,\dot{N}_{\rm ion}\sim10^{49}\,{\rm s^{-1}}$, whereas the number can increase up to 
$f_{\rm esc}\,\dot{N}_{\rm ion}\sim10^{52}\,{\rm s^{-1}}$ in the most massive halos ($\mvir > 10^{10}\,\msun$)
(Figure~\ref{fig:fesc_stat}, right panel). 
Notice, however, that this does not necessarily translate to their relative role to the reionization of the universe.
Small halos at high redshift may make a more significant contribution
to the Thompson optical depth \citep{wyithe07,shull12,kuhlen12,robertson13}.

It is noted that the recombination timescale corresponding to the mean 
density of the universe at $z\sim10$ ($n_{\rm H}\sim10^{-3}\,{\rm cm^{-3}}$) is relatively long 
($\sim$ 50--100 Myr)\footnote{Given that gas accretion is mostly filamentary 
\citep[e.g.][]{ocvirk08,dekel09,kimm11,stewart11a}, the actual density of the gas that occupies 
most of the volume in the halo is likely to be even lower than the mean density of the universe, 
and the recombination timescale could be longer.}, and thus the halo gas around a galaxy 
may be kept partially ionized even though it is irradiated by the galaxy intermittently.
Figure~\ref{fig:ex} (the second panel in the middle row) indeed shows that a large fraction of the 
IGM in the vicinity of the central galaxy is largely ionized despite the fact that instantaneous \fesc\ is low. 
Although we do not include the whole distribution 
of the ionized hydrogen inside the halo, we confirm that the halo gas between 2 kpc and 12 kpc (virial radius) 
is fully ionized apart from the small region taken by cold filamentary gas.
In fact, the volume filling fraction of the neutral hydrogen ($f_{\rm v}$) 
inside $0.2\,\rvir$ ($\sim$2.3 kpc) is found to be $\sim$ 25\% larger in the snapshot (b) ($f_{\rm v}\approx0.04$) 
than that in the snapshot (a), suggesting that dense star-forming gas plays a more important role 
in determining the escape fraction than volume-filling diffuse neutral gas. 

\begin{figure}
   \centering
            \includegraphics[width=8cm]{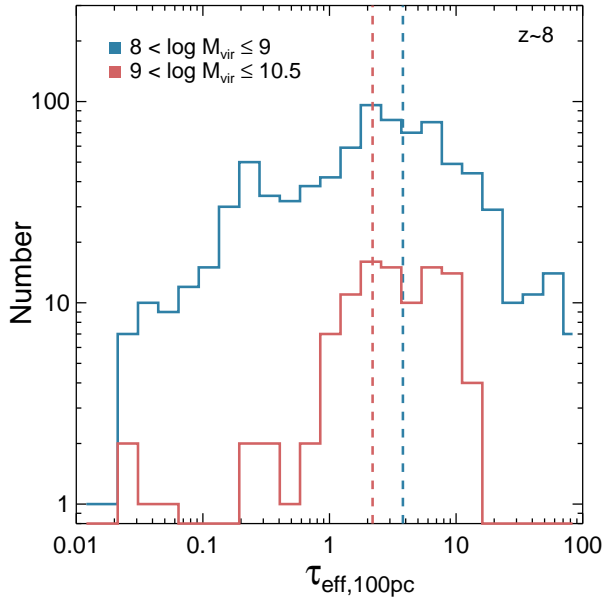}
   \caption{ Effective optical depth in the Lyman continuum ($\tau_{\rm eff}$) by the gas in the vicinity of each star 
   ($<$100 pc) in galaxies with a low escape fraction ($f_{\rm esc} < 0.1$) at $z\sim8$ from the \FRR\ run. We cast 768 rays 
   uniformly distributed across the sky for individual star particles and combine the absorption of Lyman continuum 
   by neutral hydrogen at the distance of 100 pc from each star to obtain the effective optical depth. Different color codings 
   display the distribution in different halo mass bins, as indicated in the legend.  The dashed lines indicate the 
   photon production rate-weighted average of the effective optical depth. 
  Again, we combine the results from seven consecutive snapshots to increase the sample size.
   We find that $\tau_{\rm eff,100pc}$ is generally 
   large (2 -- 4) for the galaxies with the low escape fraction, indicating that the nearby gas alone could reduce the 
   number of ionizing photons by 7 -- 45. This demonstrates that the ISM should be properly 
   resolved to better understand the escape of ionizing photons.
    }
   \label{fig:tau}
\end{figure}

Figure~\ref{fig:tau} demonstrates the importance of resolving the ISM in predicting the escape 
of ionizing photons. In order to estimate the optical depth by neutral hydrogen in the vicinity of 
each star particle ($<$ 100 pc), we spawn 768 rays per particle using the {\sc Healpix} algorithm \citep{gorski05}.
Each ray carries the spectral energy distribution determined by the age and mass of the star particle \citep{leitherer99}. 
As the ray propagates, we compute the absorption of the Lyman continuum by neutral hydrogen as, 
$F_{\rm abs} (\nu)  = F_{\rm int} (\nu) \exp{\left[-\tau_{\rm HI} (\nu)\right]}$,
where  $\tau_{\rm HI}$ ($=N_{\rm HI} \sigma_{\rm HI}$) is the optical depth and $\sigma_{\rm HI}$ is the hydrogen ionization
cross section \citep{osterbrock06}
We then combine the attenuated spectral energy distributions propagated out to 100 pc from each star particle, 
and measure the remaining number of ionizing photons ($N_{\rm ion,tot}^{\rm final}$) per galaxy. 
This is compared with the initial number of ionizing photons ($N_{\rm ion,tot}^{\rm int}$) to obtain the effective 
optical depth as $\tau_{\rm eff, 100pc} \equiv \ln \left(N_{\rm ion,tot}^{\rm int} / N_{\rm ion,tot}^{\rm final} \right)$.
Figure~\ref{fig:tau} shows the distribution of the effective optical depth by the nearby gas for the galaxies with a 
low escape fraction ($\fesc < 0.1$) at $z\sim8$. We find that $\tau_{\rm eff,100pc}$ shows a wide distribution ranging from 
0.01 to $\sim$ 100, with the photon production rate-weighted averages of $\tau_{\rm eff,100pc}=$ 3.8 and 1.9 for less 
($10^8 < \mvir \le 10^9\,\msun$) and more massive ($10^9 < \mvir \le 10^{10.5}\,\msun$) halo groups, respectively. 
This indicates that the number of escaping photons is reduced by a factor of $7-45$ due to the gas near young stars 
in galaxies with the small \fesc.  In this regard, one may find it reconcilable that 
results from cosmological simulations with limited resolutions \citep[e.g.,][]{fujita03,razoumov10,yajima11} 
often give discrepant results.

To summarize, we find that there is a time delay between the peak of star formation activity and the escape fraction
due to the delay in the onset of effective feedback processes that can blow birth clouds away. 
Because of the delay, only 11.4 \% of the ionizing photons could escape from their host halos 
when photon production rate-averaged over all halos at different redshifts, despite the fact that 
the instantaneous \fesc\ could reach a very high value temporarily. Halos of different masses 
($8\le \log \mvir\le10.5$) contribute comparably per logarithmic mass interval to reionization, and 
a photon production rate-averaged escape fraction ($\fescg(t)$) shows a weak dependence on redshift 
in the range examined \citep[c.f.,][]{kuhlen12}.

\begin{figure}
   \centering
   \includegraphics[width=8.6cm]{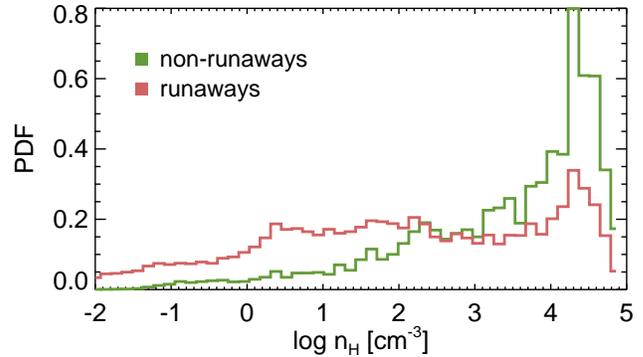}
   \caption{Difference in environment where runaway and non-runaway stars younger than 5 Myr are located.
   Approximately $2\times10^5$ stars from the most massive galaxy at $z=7$ are used to plot the histograms. 
   It can be seen that runaway stars tend to be located in less dense regions than non-runaway stars.
 }
   \label{fig:nH_runaway}
\end{figure}

\begin{figure}
   \centering
   \includegraphics[width=7.5cm]{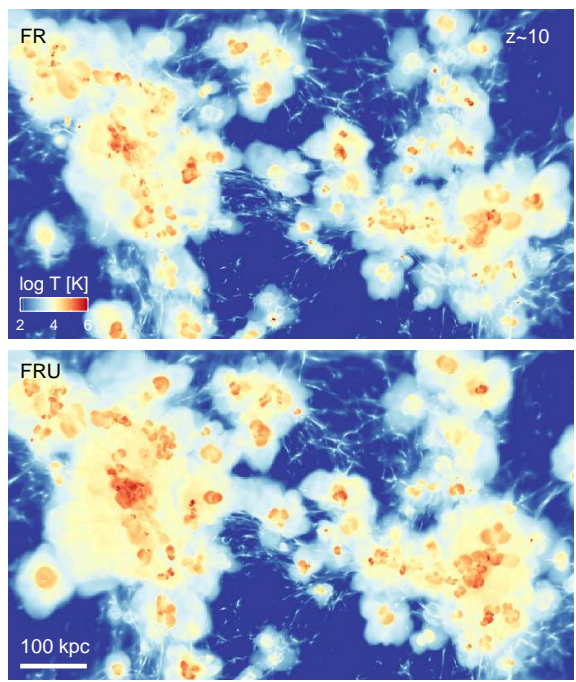}
   \caption{Comparison of the temperature distribution in the run without (top, \FRR) and with 
   runaway OB stars (bottom, \FRU) at $z=10.2$. The white bar measures 100 kpc (proper).
   The \FRU\ run shows bigger hot bubbles (30\%) with $T\ge10^5\,K$ than the \FRR\ run,
   suggesting that runway OB stars affect the regulation of star formation.
 }
   \label{fig:tem}
\end{figure}

\subsection{Escape Fraction Enhanced by Runaway OB Stars}

\begin{figure*}
   \centering
   \includegraphics[width=8.1cm]{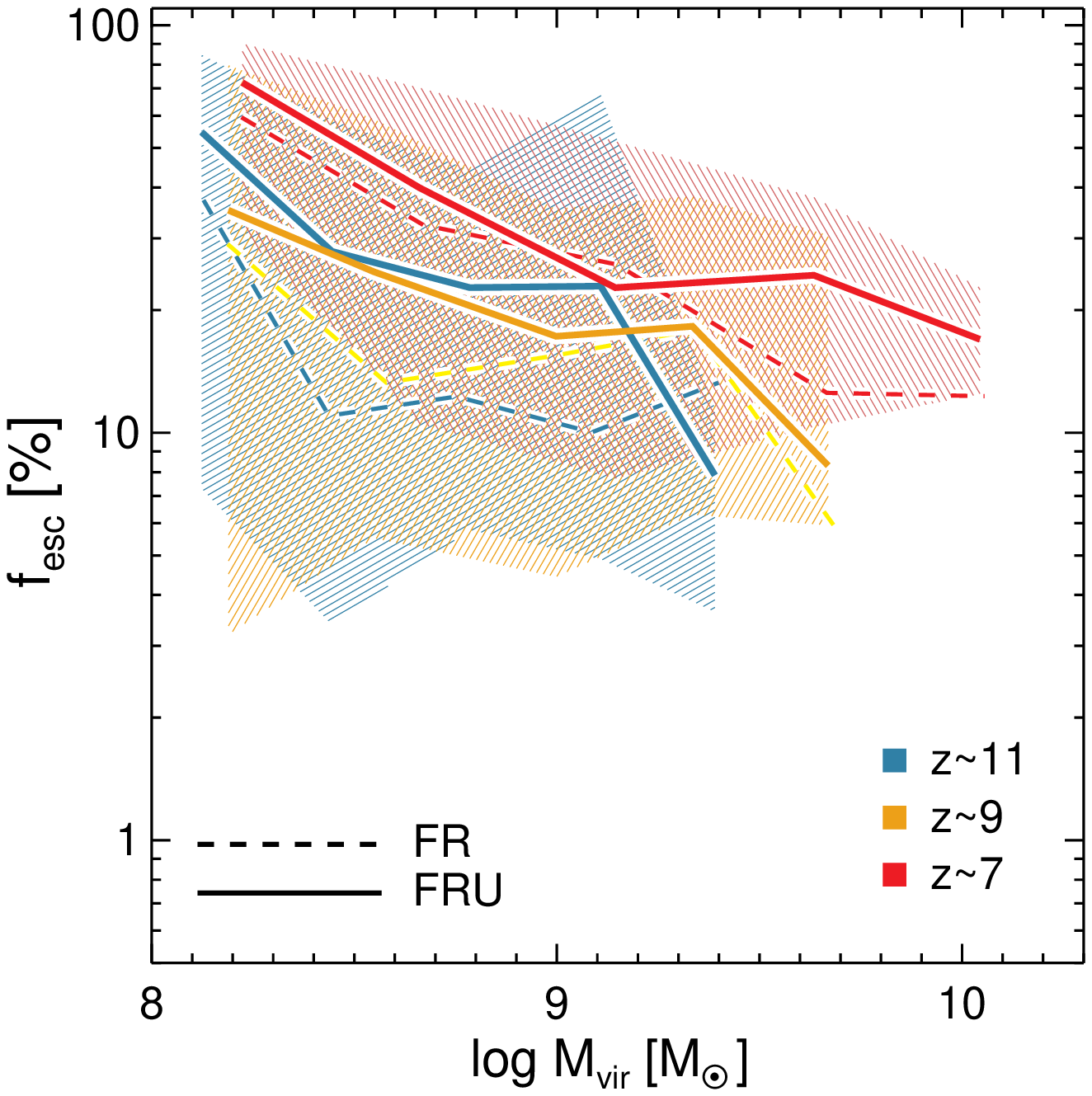}
   \includegraphics[width=8.5cm]{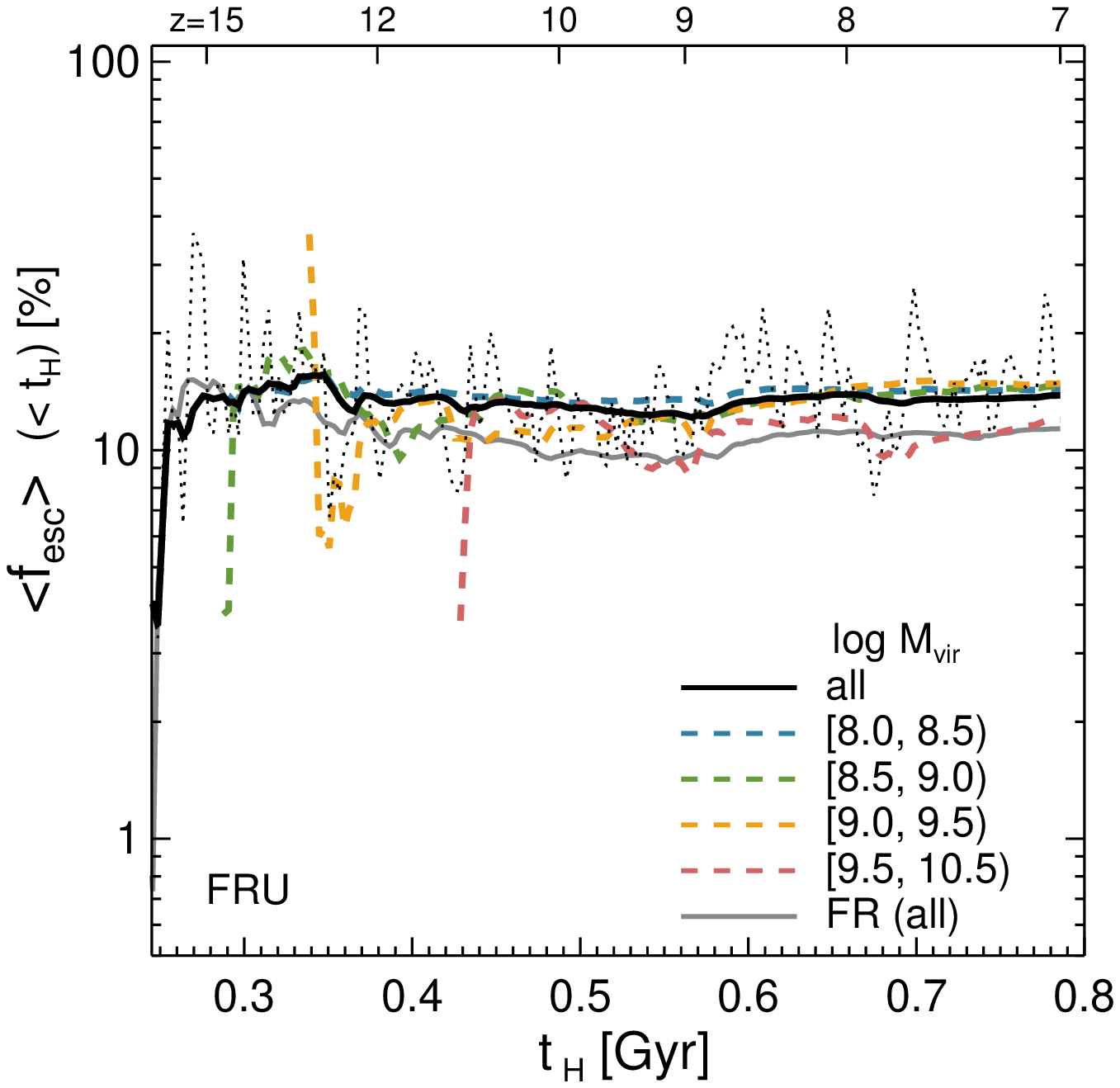}
   \caption{Impact of the inclusion of runaway OB stars on the escape fraction. {\it Left:} Instantaneous escape fraction measured 
   at the virial radius. Different color codings display different redshifts, as indicated in the legend.
   The median \fesc\ from the \FRU\ run (with runaway OB stars) and the \FRR\ run are shown as solid and dotted lines, respectively. 
   The shaded regions mark the interquartile range of \fesc\ from the \FRU\ run. It can be seen that runaway OB stars tend to 
   increase the escape probability of ionizing photons. 
 {\it Right:}  Photon production rate-weighted escape fraction, $\fescg$, 
 averaged over the 
 age of the universe ($t_{\rm H}$). The black lines include the whole sample of the simulation, 
 while the results in different halo mass bins are presented as dashed lines with different colors. 
 The solid and dashed lines show the time-averaged $\fescg$, while the dotted line 
 shows a measurement of $\fescg$ for all halos at each snapshot. 
 The time-averaged escape fraction of $\fescg$ measured at $z=7$ 
 is 13.8\% in this simulation. We find that the inclusion of runaway OB stars 
 increases the escape of ionizing photons by 22\% by $z=7$, compared with that from the \FRR\ run.
 }
   \label{fig:fesc_runaway}
\end{figure*}

Ionizing photons can not only escape from their birth clouds by destroying them through feedback processes,
but also emerge from runaway OB stars displaced from the birth clouds.
If we take the typical velocity of the runaway OB stars  
$\sim\,40\,{\rm km\,s^{-1}}$ \citep{stone91,hoogerwerf01,tetzlaff11}, 
they could travel a distance of $\sim$ 200 pc in 5 Myr. 
\citet{conroy12} examined the possible ramification of the inclusion of the runaway OB stars 
using a simple analytic formulation, and concluded  that \fesc\ can be enhanced by a factor of 
up to 4.5 from $\fesc\approx0.02-0.04$ to $\fesc\approx0.06-0.18$ 
in halos of mass $10^8 \lesssim \mvir \lesssim 10^{9}\,\msun$.
Given the complexity of the ISM dynamics \citep[e.g.][]{mckee07}, 
it would seem prudent to examine this issue in greater details in realistic environments.
To do so, we have performed a twin cosmological simulation of 
the \FRR\ run by designating 30\% of mass in each stellar particle as a separate runaway particle 
and dynamically follow their motion.

Figure~\ref{fig:nH_runaway} shows an example of the difference in environment 
between runaway and non-runaway particles in a galaxy in a $3\times10^{10}\,\msun$ halo at $z=7$. 
At this redshift, the central galaxy shows $\fesc=0.14$.
The average hydrogen number density for runaways younger than 5 Myr ($n_{\rm H}\sim130\,{\rm cm^{-3}}$) is found to be 
roughly 20 times smaller than that of non-runaways ($n_{\rm H}\sim3000\,{\rm cm^{-3}}$).
Given that these stars will explode in the next 5--10 Myrs, the fact that the local density of some runaway OB stars 
is smaller than non-runaways suggests that the impact from SN explosions will be enhanced.
Indeed, we find that the stellar mass of the galaxies in halos of mass $\mvir\gtrsim10^9\,\msun$ is smaller by a factor 
of 1.7 on average, compared with that from the \FRR\ run (see Figure~\ref{fig:mstar}). 
For galaxies in smaller halos, there is no clear hint that the runaway OB stars help suppress the star formation.
This is partly because runaway OB stars can not only provide energy but also distribute metals more efficiently, 
which can increase the cooling rate in halos. Comparison of the temperature distribution between the 
two runs further substantiates the claim that runaway OB stars help regulate the star formation (Figure~\ref{fig:tem}). 
The volume of $T\ge10^{5}\,{\rm K}$ gas inside the zoomed-in region in the \FRU\ run ($\approx$ 7 kpc$^3$, physical)
is 30\% larger than that in the \FRR\ run.

The left panel in Figure~\ref{fig:fesc_runaway} shows the instantaneous \fesc\ measured 
at three different redshifts from the \FRU\ run. Again, less massive galaxies tend to exhibit a 
higher \fesc, which can be attributed to the fact that star formation in smaller halos is more easily affected 
by the energetic explosions. As expected, the inclusion of the runaway OB stars 
increases the instantaneous escape fraction on average. The photon production rate-weighted average 
of \fesc\ (right panel in Figure~\ref{fig:fesc_runaway}) shows this more clearly. In our fiducial run (\FRR), 11.4\% of 
the ionizing photons produced escaped from the halos of mass $\mvir\ge10^8\,\msun$ at $z\ge7$. 
On the other hand, the \FRU\ run yields higher $\fescg$ of 13.8\%, 
which is enhanced by 22\% compared with that of the \FRR\ run.
Although this increase is not as large as claimed in \citet{conroy12}, 
the contribution from the runaway OB stars is certainly significant. 
Similarly as in the \FRR\ run, no clear dependence of $\fescg$ on halo mass is found.

It is interesting to discuss possible origins of the significantly different enhancement in the escape fraction
due to runaway OB stars found in our simulations compared with the estimate by \citet{conroy12}. 
First, while their model predicts \fesc\ of non-runaways to be about 2--4\% in halos of mass 
$10^8 \le \mvir \le 10^9 \, \msun$, we find that the self-regulation of star formation via SN explosions 
leads to a high escape of $\sim$ 10\% in our fiducial model (\FRR). 
Second, while their model finds that runaway OB stars are found to have high $\fesc$ (=30--80\%), 
our results imply that the mean escape fraction of ionizing photons from runaway OB stars
is about $20\%$ ($11.4\%\times 70\% + {\it 20\%}\times 30\%\approx13.8\%$).
We also make a more elaborate estimate as follows.
We measure the optical depth in the Lyman continuum for the gas inside each halo along 768 sightlines 
per star particle, and combine the attenuated spectral energy distributions. These are used to count 
the number of hydrogen ionizing photons for runaways and non-runaways separately. 
We find that the relative contribution from the runaways to the total number of escaping photons is
comparable with that of the non-runaways. Considering that the runaway particle is assumed to explain only 30\% of 
all the OB stars, the net $\fesc$ for the runaways can be estimated to be roughly 23\% ($=13.8\%/2/0.3$). 
This is twice higher chance of escaping than the non-runaways, but much smaller than computed in 
the analytic model. If the escape fraction of non-runaway OB stars were 2\% in our simulations, 
the total escape fraction would become $2\%\times 70\% + 23\%\times 30\%=8.3\%$, 
corresponding to an increase of a factor of 4.2.
It is thus clear that most of the discrepancies arise in a large part due to different escape fraction values 
for non-runaway OB stars and also due to different escape fraction values for runaway OB stars.

\begin{figure}
   \centering
            \includegraphics[width=8.6cm]{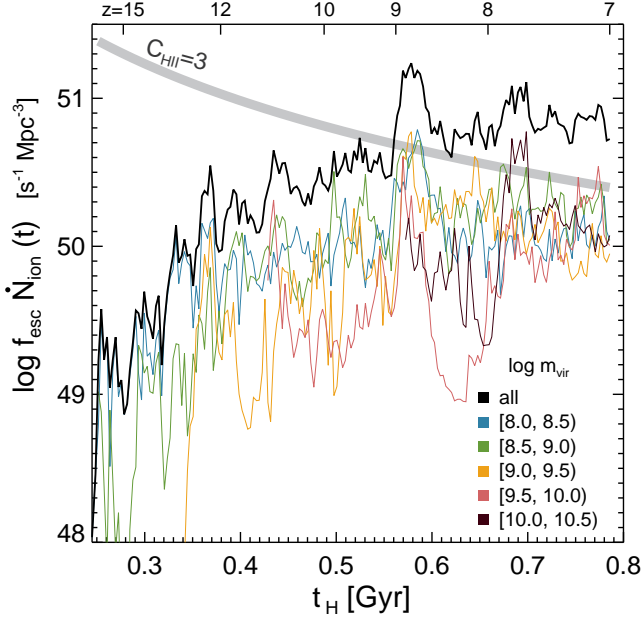}
   \caption{Balance between the ionizing photons escaping from the dark matter halo and the recombination rate 
   in the \FRU\ run.
   The thick grey line shows the balance condition when the clumping of $C_{\rm HII}=3$ is used. 
   Enough photons to keep the universe ionized  escape from the halo after $z\sim8$.
   }
   \label{fig:budget}
\end{figure}

Although $\fescg$ is 22\% larger in the \FRU\ run than \FRR, the cumulative number of photons escaped in halos 
with $\mvir\ge10^8\,\msun$ by $z=7$ ($N_{\rm ion}\approx1.3\times10^{69}$) is found to be similar to that of the \FRR\ run ($N_{\rm ion}\approx1.6\times10^{69}$).
This is because star formation is suppressed in relatively massive halos ($\mvir \ge 10^9\,\mvir$).

\begin{figure}
   \centering
            \includegraphics[width=8.6cm]{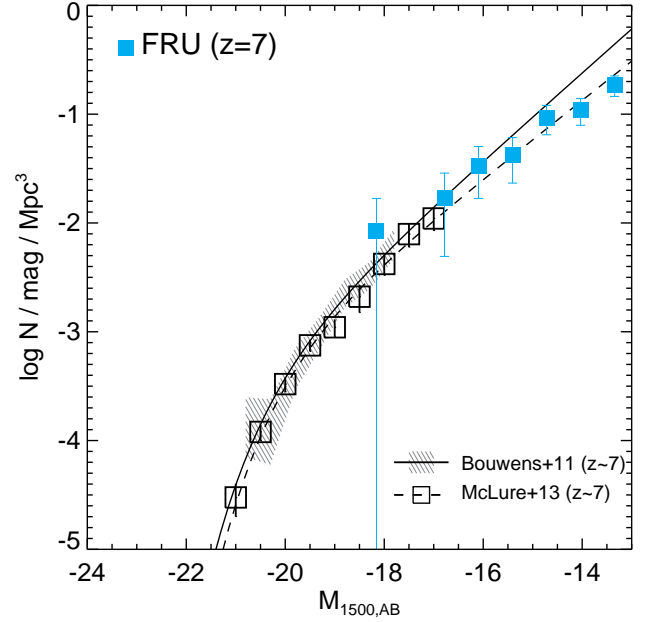}
   \caption{Rest-frame ultraviolet luminosity function from the \FRU\ run at $z=7$.  Error bars denote the Poissonian error.
   Observational data from \citet{bouwens11a} and \citet{mclure13} are shown as the shaded region 
   and empty squares, respectively. Also included as solid and dashed lines are the Schechter fits to the 
   data provided in these studies.
   }
   \label{fig:uvlf}
\end{figure}

One question is whether or not enough photons escape to keep the 
universe at $z\sim7$ ionized.  The critical photon rate density that can balance the recombination of ionized hydrogen is 
\begin{equation}
\dot{n}_{\rm ion}^{\rm crit}  = \alpha_{\rm B} \, n_e \, n_{\rm HII} \simeq 10^{47.2} C_{\rm HII} (1+z)^3 \, {\rm [s^{-1}\,Mpc^{-3}]},
\end{equation}
where $\alpha_B$ is the case B recombination coefficient, 
$n_e$ is the number density of electron, $n_{\rm HII}$ is 
the number density of ionized hydrogen, and $C_{\rm HII} \equiv \left<n_{\rm HII}^2\right>/\left<n_{\rm HII}\right>^2$ is the 
clumping factor of ionized gas. For a  choice of the clumping factor $C_{\rm HII}\sim3$ \citep{pawlik09,raicevic11} and 
the temperature $T=20000K$, $\dot{n}_{\rm ion}^{\rm crit} = 10^{50.4}\,[(1+z)/8]^3  \, {\rm s^{-1}\,Mpc^{-3}}$.
Figure~\ref{fig:budget} shows that the escaped photons in \FRU\ can balance the recombination at $z\le 9$. 
We find that the photon rate density at $z\sim7$ is $\dot{n}_{\rm ion}=10^{50.7-50.9}   \, {\rm s^{-1}\,Mpc^{-3}}$,
consistent with observational findings. \citet{ouchi09} estimated the ionizing photon density to be 
$\log \dot{n}_{\rm ion} \simeq 49.8 - 50.3$ by integrating the UV luminosity function (UVLF) down to $M_{\rm UV}=-18$ (lower) 
or $L=0$ (upper estimate) with a slope of $\alpha=-1.72$ at $z\sim7$ with $\fesc=20\%$. 
If the slope found in the more recent literature \citep{mclure13}, $\alpha=-1.90$, 
is used,  the maximum photon rate density derived would increase to $\log \dot{n}_{\rm ion} \simeq 50.8$,
which is in agreement with our estimation. Note that the photons escaping from halos of mass 
$\mvir\ge10^8\,\msun$ account for more than 90\% of the total escaping photons 
if the baryon-to-star conversion efficiency derived in our simulation 
is extrapolated to smaller halos ($\mvir<10^{8.5}\,\msun$, see below), 
and hence our results should be compared with the maximum photon rate density.
Given that their chosen \fescg\  is closed to what our simulation yields (13.8\%), 
the agreement implies that SFRs of the galaxies are well reproduced in our simulation. Indeed, 
we find that our simulated UVLF measured at 1500\AA\ (rest-frame)  shows excellent agreement with 
the LF with the slope of $\alpha=-1.90$ \citep{mclure13} down to $M_{\rm 1500}=-13$ (Figure~\ref{fig:uvlf}).
Here we neglect the effect of dust extinction, as the galaxies in our sample are very metal-poor 
($Z_{\rm star}\lesssim10^{-3}$).

\begin{table} 
\caption{Photon number-weighted $f_{\rm esc}$ at $7\le z \lesssim 15$ from the FRU run}
\centering
\begin{tabular}{@{}cc}
\hline 
$\log M_{\rm vir}$ & $\left<f_{\rm esc}\right>$       \\
\hline 
8.25     & 0.144 $\pm$ 0.038\\
8.75     & 0.146 $\pm$ 0.064\\
9.25     & 0.148 $\pm$ 0.077\\
9.75     & 0.128 $\pm$ 0.069\\
10.25   & 0.113 $\pm$ 0.079\\
\hline 
\label{table2}
\end{tabular} 
\end{table}

\begin{figure}
   \centering
            \includegraphics[width=8.6cm]{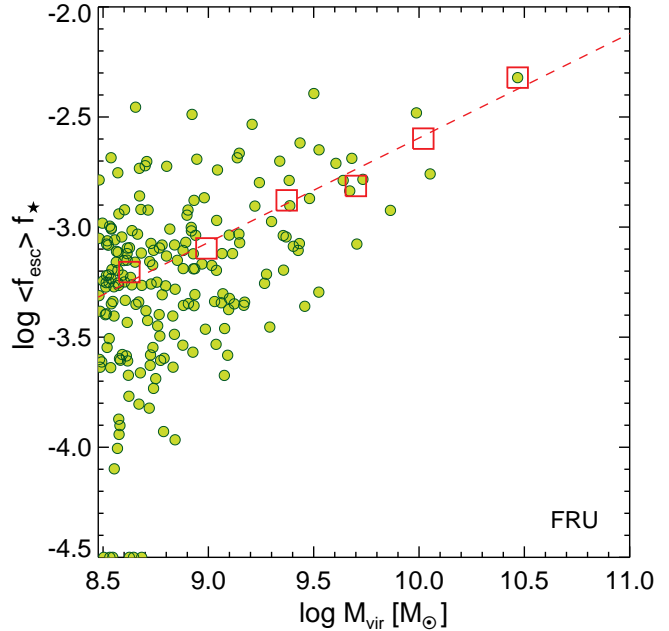}
   \caption{Product of the stellar mass fraction within the virial radius of a dark matter halo 
   ($f_\star=\Omega_{\rm m} \mstar / \Omega_{\rm b} \mvir$)  at $z=7$ 
   and halo mass-dependent photon production rate-averaged escape fraction from the cosmological simulation with 
   runaway OB stars (\FRU).
   Averages are shown as red empty squares, with the simple regression (dashed line). 
   A smaller number of photons is escaped per unit mass in smaller halos, reflecting the results that
   star formation is inefficient in the low-mass halos. 
   }
   \label{fig:cstar}
\end{figure}

In Figure~\ref{fig:cstar}, we plot the product of photon number-weighted escape fraction ($\fescg$) and 
 baryon-to-star conversion efficiency ($f_{\star}\equiv \Omega_{\rm m} M_{\rm star}/\Omega_{\rm b} M_{\rm vir}$) 
  at $z=7$. Notice that we include all stars within the virial radius of a dark matter halo 
  in this measurement. Since there is little evolution in $\fescg$ with redshift (Figure~\ref{fig:fesc_runaway}, 
  right panel), we combine $\fescg$ of the halos in the same mass 
  range at $7\le z < 20$ to obtain the mean escape fraction as a function of halo mass (Table~\ref{table2}).
  We then use a simple fit to the mean, as
  \begin{equation}
  \log \fescg (\mvir) \approx -0.510 - 0.039 \log \mvir.
  \label{fescg_fit}
  \end{equation}
We limit our fit to the sample with $\mvir\ge10^{8.5}\,\msun$, where each halo is 
 resolved with $\sim$ 2000 dark matter particles and more.
There is a trend that more massive halos contribute more to the total number of ionizing photons per mass,
which essentially reflects the fact that low-mass halos are inefficient in forming stars 
(see also Figure~\ref{fig:mstar}). The average $\fescg f_{\star}$ of different halo masses can be  
fitted with 
\begin{equation}
\log \fescg f_{\star} \approx -7.342 + 0.474\, \log \mvir,
  \label{fstar_fit}
\end{equation}
shown as the red dashed line in Figure~\ref{fig:cstar}.
We note that $\fescg f_{\star}$  becomes as low as $\sim 5\times10^{-4}$ 
in small halos ($\mvir\sim10^{8.5}\,\msun$), which is roughly 40 times smaller than the results 
from \citet{wise09} ($\fescg f_\star \approx0.02$). 
The difference can be attributed to two factors.  
First, our \fescg\ is smaller by a factor of $\sim3-4$ than that of \citet{wise09}. 
This is probably due to the fact that their cosmological runs start from the initial condition 
extracted from adiabatic simulations in which no prior star formation is included. 
Since radiative cooling and star formation are suddenly turned on at some redshift,
the gas in the halo rapidly collapses and forms too many stars in their cosmological runs.
This is likely to have resulted in stronger starbursts in the galaxies, leading to a higher escape probability. 
Second, because of the same reason, $f_{\star}$ is considerably higher in the \citet{wise09} halos than in our halos. 
For halos of masses with $\mvir\sim10^{8.5}\,\msun$, we find that $f_{\star}\approx0.003$, 
which is smaller by a factor of $\sim 10$ than those in \citet{wise09}. 
Indeed, we find fairly good agreement with the latest determination of $\fescg f_{\star}$ in 
halos of $\mvir\sim10^{8.5}$ by \citet{wise14},
who model star formation self-consistently in their cosmological radiation hydrodynamics simulations.

It is worth mentioning that adopting high spatial resolution (or gravitational softening length) 
is important to accurately predict the escape fraction. If the resolution is not high enough to capture 
the rapid collapse of gas clouds, the resulting star formation histories would become less episodic, 
leading to a longer time delay between the peak of star formation and escape fraction. 
This in turn would reduce the fraction of escaping photons.  To examine this issue, 
we run two additional simulations with the identical initial condition and other parameters, 
but with one less or more level of refinement, corresponding to 8.5 pc or 2.1 pc (physical) resolution, respectively.
We find that the run with the lower resolution yields a factor of two smaller mean escape fraction at z=9 
($\fescg=7.6\%$, see Appendix). On the contrary, higher resolution run exhibits a comparable mean 
escape fraction of $\fescg=13.9\%$ at $z=10$, suggesting that the results are reasonably converged 
for the parameters used in the \FRU\ run.

\section{Discussion}

Recent studies show that the escape fraction should be larger than 20\% to re-ionize the 
universe by $z=6$ matching the Thomson optical depth inferred from the CMB 
\citep{kuhlen12,shull12,robertson13}. This can be obtained by numerically solving the simple differential 
equation for the \hii\ bubble
\begin{equation}
\frac{d Q_{\rm HII} }{dt} = \frac{\dot{n}_{\rm ion}}{\left<n_{\rm H}\right>} - \frac{Q_{\rm HII}}{t_{\rm rec}(C_{\rm HII})},
\end{equation}
where $Q_{\rm HII}$ is the volume filling fraction of the bubble, $\left< n_{\rm H}\right>$ is the comoving mean density 
of the universe, and 
$t_{\rm rec} (C_{\rm HII})= \left[ C_{\rm HII}\,\alpha_{\rm B}(T)\, f_e\, \left<n_{\rm H}\right> \,(1+z)^3\right]^{-1}$ is the 
recombination timescale for a given clumping factor and temperature. Here $f_e$ is a correction factor that accounts 
for the additional contribution of singly ($z>4$) or doubly ($z<4$) ionized helium to 
the number density of electron \citep[e.g.,][]{kuhlen12}. 
We adopt a redshift-dependent clumping factor of $C_{\rm HII} = 1 + \exp(-0.28\, z +3.59)$ at $z\ge10$ or
$C_{\rm HII} = 3.2$ at $z<10$ following \citet{pawlik09}.
Once $Q_{\rm HII}$ is determined, the Thomson optical depth 
as a function of redshift can be calculated as 
\begin{equation}
\tau_e (z)= \int_0^z c \left<n_{\rm H}\right>\,\sigma_T\,f_e\,Q_{\rm HII}(z') \frac{(1+z')^2 dz'}{H(z')},
\end{equation}
where $\sigma_T$ is the Thomson electron cross section, and $H(z)$ is the Hubble parameter.
We follow the exercise by using the ionizing photon density from Figure~\ref{fig:budget} to examine 
whether our models provide a reasonable explanation for the reionization history.
For $\dot{n}_{\rm ion}$ at  $z<7$, we extrapolate based on the simple fit to the results in Figure~\ref{fig:budget}. 
This simple experiment indicates that the universe can be re-ionized by $z=7.25$. 
However, the evolution of the photon density from the \FRU\ run predicts a smaller volume filling fraction of the \hii\ bubble
at $z=10$ ($Q_{\rm HII}=12\%$), compared with other analytic models \citep[$Q_{\rm HII}\gtrsim20\%$, e.g.][]{shull12} 
that could reproduce the CMB measurement \citep[$\tau_e\sim0.09$,][]{komatsu11}. Consequently,  the FRU run yields 
the Thomson optical depth of $\tau_e=0.065$, which is consistent only within 2$\sigma$ with the 
CMB measurement. This implies that more ionizing photons are required to escape from halos at high redshift 
to explain the reionization history of the Universe.

\begin{figure}
   \centering
            \includegraphics[width=8.6cm]{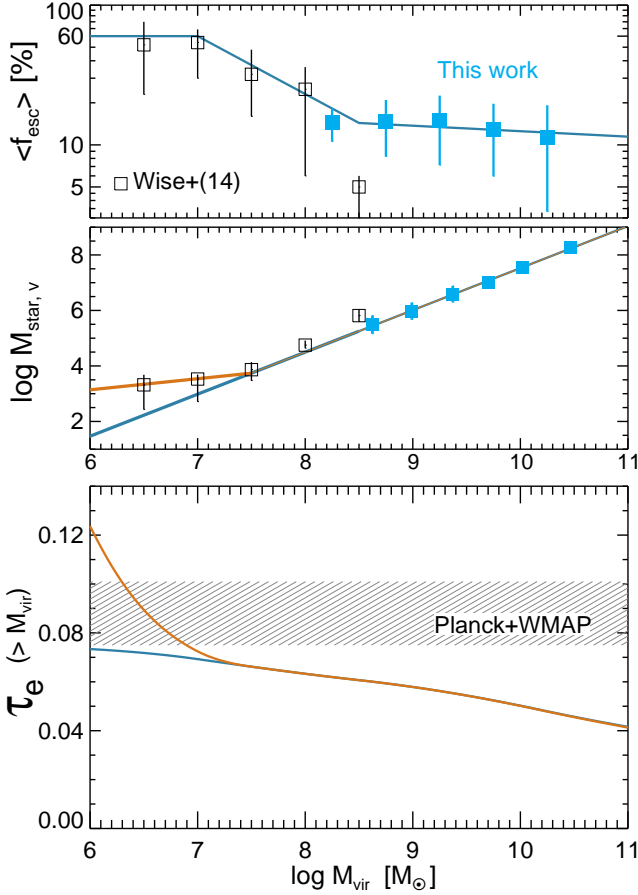}
   \caption{ Importance of the dwarf galaxy population to the Thomson optical depth measurement 
   in semi-analytic calculations. The top and middle panels show the escape fraction and 
   stellar mass inside the virial radius of a halo as a function of halo mass, respectively, 
   which are used to compute the optical depth (the bottom panel). The measurements from 
   our radiation cosmological simulations with runaway stars (\FRU) are shown as 
   blue filled squares with the standard deviations. Empty squares with error 
   bars are the results from \citet{wise14}.    The optical depth is obtained 
   by taking into account the escaping ionizing photons from halos more massive than $\mvir$.
   We neglect the contribution from rare massive halos with $\mvir>10^{12}\,\msun$. 
   Different colors in the bottom panel corresponds to the results with different assumptions on 
   the stellar-to-halo mass relation for minihalos, as indicated in the middle panel.
   The shaded region denotes the Thomson optical depth inferred from the Planck+WMAP
   measurements.
   }
   \label{fig:tau_es}
\end{figure}

The deficiency of ionizing photons may in part be attributed to the fact that 
our simulations cannot resolve the collapse of small-mass halos ($\mvir\lesssim10^8\,\msun$)
due to finite mass resolution. \citet{Paardekooper13} argue that reionization is driven by 
dwarf-sized halos of masses  $\mvir=10^7-10^8\,\msun$ with high $\fescg$ of $\approx$0.4--0.9.
Similarly, \citet{wise14} find that the ionizing photons from the minihalos with 
$\mvir=10^{6.25}-10^{8.25}\,\msun$ is crucial at reproducing the Thompson optical 
depth from the CMB measurements. In order to examine the importance of the minihalos 
in light of our new results, we estimate the optical depth as a function of the minimum halo mass
that can contribute to reionization. To do so, we use the theoretical halo mass functions at different 
redshifts \citep{jenkins01},  convolved with the baryon-to-star conversion efficiency measured at $z=7$ from 
the \FRU\ run for $\mvir\ge10^{7.5}\,\msun$ and \citet{wise14} for $\mvir<10^{7.5}\,\msun$ 
(orange line in the middle panel of Figure~\ref{fig:tau_es}), to derive the increase in the stellar mass density 
with redshift. The number of escaping ionizing photons is then calculated by multiplying the number of 
photons produced with the halo mass-dependent escape fraction based on our results and \citet{wise14}, 
as (Figure~\ref{fig:tau_es}, top panel)
\begin{equation}
\log \fescg=\left\{
\begin{array}{ll}
 -0.51 - 0.039\,\log \mvir &  (\log \mvir \ge 8.5) \\
 2.669 - 0.413\,\log \mvir &  (7 \le \log \mvir < 8.5) \\
 -0.222 &  (\log \mvir < 7)\\
\end{array} 
\right. .
\end{equation}
We neglect the contribution from rare massive halos with $\mvir>10^{12}\msun$.
Figure~\ref{fig:tau_es} (orange line, bottom panel) shows that 
the minihalos of $\mvir<10^7\,\msun$ can indeed provide enough photons 
to match $\tau_e$ inferred from the CMB measurement. 
While the ionizing photons from $\mvir>10^7\,\msun$ only gives $\tau_e=0.072$,
the additional photons arising from the minihalos augment the optical depth to 0.122.
However, we note that this sensitively depends on the assumption
on the baryon-to-star conversion efficiency in the minihalos. For example, when the stellar 
mass-halo mass relation found in the \FRU\ is extrapolated to the minihalos 
(blue line in the bottom panel), the optical depth for the entire halos is only $\tau_e=0.073$.
Given that these minihalos would host a handful of star particles with $m_{\rm star}\sim10^2-10^3\,\msun$ 
in current numerical simulations, it is unclear how the mass resolution affects the conversion efficiency, 
and further investigations on star formation in the minihalos will be useful to better understand 
their relative role to the total ionizing budget.

In our simulation, we approximate that massive stars ($M>8\msun$) evolve off and 
explode after 10 Myr. We note that this is roughly the timescale 
of the delay between the peak of star formation and escape fraction.
In reality, the SN can emerge as early as $\sim$ 3 Myr for a simple population \citep{schaller92}. 
Stellar winds, photo-ionization, and radiation pressure acting on electron and dust can come into play even earlier. 
\citet{walch12} claims that a $10^4\,\msun$ molecular cloud of the radius 6.4 pc can be dispersed 
on a 1-2 Myr timescale by the overpressure of \hii\ regions.
Moreover, it is also plausible that the ionization front instabilities may lead to the higher escape probability 
of ionizing photons \citep{whalen08b}.
If these mechanisms played a role in shaping the evolution of individual molecular clouds, the escape fraction 
measured in our simulations would have been higher than 14\%. 
In this regard, our photon number-weighted mean is likely to represent the minimum escape of ionizing photons. 
When a higher \fescg\ of 30\% is assumed for the star formation history in the \FRU\ run, 
dark matter halos of $\mvir>10^8\,\msun$ alone can achieve $\tau_e=0.076$,
suggesting that a more precise determination of the escape fraction is as equally important 
as resolving ultra-faint galaxies with $M_{\rm 1500} > -13$. 
Future studies focusing on the interplay between the feedback processes will shed more light 
on the reionization history of the Universe.

\section{Conclusions}

The escape fraction of hydrogen ionizing photons is a critical ingredient in the theory of reionization. 
Despite its importance, only a handful of studies examined the escape fraction ($\fesc$)
of high-$z$ galaxies in a cosmological context \citep{wise09,razoumov10,yajima11,Paardekooper13,wise14}.
To better understand the physics behind the escape of ionizing photons and quantify \fesc, 
we have carried out two zoomed-in cosmological radiation hydrodynamics simulations of 
$3.8\times4.8\times9.6$ Mpc$^3$ box (comoving) with 
the \ramses\ code  \citep{teyssier02,rosdahl13} with high spatial ($\sim$ 4 pc, physical) and 
stellar mass resolution of 49 $\msun$.
Because energy-based feedback from SN explosions suffers from the artificial
radiative cooling if the cooling length is under-resolved, we have implemented a new 
mechanical feedback scheme that can approximate all stages of a SN explosion 
from the free expansion to snowplow phase. 
With the physically based feedback model, 
we have investigated the connection between the regulation of star formation and 
corresponding evolution of the escape of ionizing photons.  
We have also explored the relative importance of runaway OB stars to the escape fraction 
by comparing the twin simulations with (\FRU) and without (\FRR) runaways.
Our findings can be summarized as follows.

\begin{enumerate}

\item When a dense cloud begins to form a cluster of stars, the escape fraction is negligible. 
As energetic explosions by massive stars follow after $\sim$ 10 Myr, it blows the star forming gas away,
increasing the {\it instantaneous} escape fraction (\fesc) to \gtrsim10\%. Although \fesc\ is kept high in this phase,
subsequent star formation is markedly suppressed, 
and only a small number of photons escapes from their host dark matter halo (Figure~\ref{fig:ex}). 
This time delay between the peak of star formation and the escape fraction is crucial in predicting 
the actual escape probability of ionizing photons. While the instantaneous \fesc\ can easily 
attain $\gtrsim30\%$ in halos of mass $\mvir \ge 10^8\,\msun$ on average (Figure~\ref{fig:fesc_stat}), 
the photon number-weighted mean of the escape fraction (\fescg) is found to be 11.4\% (Figure~\ref{fig:fesc_wei}).

\item  \fesc\ tends to be higher in less massive halos and at lower redshift for a give halo mass (Figure~\ref{fig:fesc_stat}).
This is essentially because less dense and smaller galaxies are more susceptible to SN explosions.
However, the photon production rate-averaged escape fractions show no clear dependence 
on halo mass and redshift, again implying that the interplay between star formation and the delay in the onset of 
negative feedback is more important in determining the actual escape probability. 

\item Absorption of ionizing photons by neutral hydrogen in the ISM is significant (Figure~\ref{fig:tau}). For galaxies 
with a low escape fraction ($\fesc<10\%$), the effective optical depth by the gas within 100 pc 
from each young star particles is found to be $\tau_{\rm eff,100pc}\sim 1.9-3.8$ at $z\sim8$.
The nearby neutral gas alone can reduce the number of ionizing photons by 7--45 in this case, 
demonstrating the importance of properly resolving the ISM to predict a more accurate escape fraction.

\item Our physically based SN feedback effectively regulates star formation. 
Only 0.1\% to 10\% of the baryons are converted into stars in galaxies at $z=7$ (Figure~\ref{fig:mstar}).
The energetic explosions sometimes completely shut down star formation when galaxies are small.
The baryon-to-star conversion ratio is smaller in less massive halos. 
Consequently, halos of different masses contribute comparably to the total number 
of ionizing photons escaped by $z=7$ (Figure~\ref{fig:fesc_wei}).

\item Inclusion of runaway OB stars increases the escape fraction to $\fescg=13.8\%$ from 11.4\% 
(Figure~\ref{fig:fesc_runaway}). Since the runaway OB stars tend to move to lower density regions, 
photons from them have a higher chance of escaping. Moreover, as the runaway OB stars explode in a less 
dense medium, feedback from SNe becomes more effective, resulting in reduced star formation in halos $\mvir \ge 10^9\,\msun$, 
compared with the \FRR\ run. Because of the balance between the increase in \fescg\ and the decrease 
in star formation, the total number of ionizing photons escaped by $z=7$ is found to be comparable in
the two runs.

\item 
A sufficient amount of photons escape from the dark 
matter halos with $\mvir\ge10^8\,\msun$ to keep the universe ionized at $z\le9$. 
The simulated UV luminosity function with a faint end slope of -1.9 is consistent with observations.
\end{enumerate}

\acknowledgements{
We thank an anonymous referee for constructive suggestions that improved this paper.
We are grateful to Julien Devriendt, Sam Geen, Chang-Goo Kim, Eve Ostriker,  Adrianne Slyz,
and John Wise for insightful discussions.
Special thanks go to Romain Teyssier and Joakim Rosdahl for sharing their radiation 
hydrodynamics code with us. Computing resources were provided in part by the NASA High-
End Computing (HEC) Program through the NASA Advanced
Supercomputing (NAS) Division at Ames Research Center and in part by 
Horizon-UK program through DiRAC-2 facilities.  
The research is supported by NSF grant AST-1108700 
and NASA grant NNX12AF91G.
}

\newpage

\appendix{

\section{A New Physical Scheme For Supernova Feedback}
It is well established that feedback from SN explosions is crucial to understanding many 
aspects of galaxy evolution \citep[e.g.,][]{dekel86}. 
Detailed implementation of SN feedback has progressed over time.
Early works have included it by depositing internal energy into the parent cell of the SN particle,
but the energy is found to be radiated away quickly, making the feedback ineffective \citep[e.g.][]{katz92,abadi03,slyz05,hummels12}. 
To alleviate the problem, \citet{cen06} distributed the thermal energy over 27 neighboring cells 
weighted by specific volume, which has the virtue of being able to mimic 
propagation of an explosion that is preferentially channelled into more diffuse regions.
Some groups have adopted unconventional schemes
to regulate star formation and gas dynamics.
These include increasing the total energy from the SN \citep{thacker00}, making the SN event more 
episodic and stronger \citep{scannapieco06,dalla-vecchia12}, decoupling hydrodynamic 
interactions of super-wind particles with the ambient medium \citep{oppenheimer06},
turning off gas cooling after deposition of SN energy for a significant period of time \citep{mori97,stinson06,governato07},
or disabling cooling when the turbulence measured locally is significant  \citep{teyssier13}.  
Alternatively, the explosion can be provided in kinetic form \citep{navarro93,dubois08,dalla-vecchia08}.
This is not immune to the artificial cooling problem, because the kinetic energy is 
converted into heat through shocks immediately after launching given the typical high Mach number,
ambient density, and limited numerical resolution.

So far there is no satisfactory scheme for modeling the SN feedback that is robust and does not strongly 
depend on simulation resolution. We have devised a new physical scheme that is reasonably accurate.
Before going into details, it is useful to gain a basic physical understanding of the Sedov explosion.
The Sedov explosion consists of four stages.
In the first stage, the SN ejecta sweep up an insignificant amount of mass compared with the initial ejecta mass,
and both energy and momentum are conserved.
In the second stage, the swept-up ISM mass is now comparable to or exceeds the initial ejecta mass so the explosion
now enters a self-similar phase. Cooling has not set in so energy is conserved up to this stage.
As a result, the radial outward momentum increases as the square root of the total shell mass.
The third stage is the cooling phase, which is normally very brief in an isobaric gas.
The last stage is called the snow-plow phase, when the total linear radial outward momentum is conserved.
In essence, the problem of excessive cooling in some simulations
is a result of not giving the surrounding gas an adequate amount of momentum that 
is commensurate with the stage it is supposed to be in, given the physical conditions.
For example, in some cosmological simulations with limited resolution,
a single cell is already larger than the expected radius at which the snow-plow phase commences.
In this case, a physically correct way would be to deposit the maximum expected momentum, as at the 
end of the cooling phase (i.e., the end of the third stage), to the surrounding cells. If one instead 
deposits thermal energy with no momentum or a combination of thermal energy and
an inadequate amount of momentum, the final momentum would fall short of the true expected momentum.
We now describe our new physical scheme in detail.

The momentum input from SN explosions is modeled as
\begin{equation}
dp=\left\{
\begin{array}{cc}
 \sqrt{2 \chi(\Omega) M_{\rm ej}\,f_e\,E_{\rm SN}}\times  \frac{d\Omega}{4\pi} &~~~~  \left[ \chi(\Omega) \le \chi_{\rm tr}(\Omega)\right]\\
 & \\
 p_{\rm SN}(E_{\rm SN},\,n_{\rm H}, Z) \times \frac{d\Omega}{4\pi} & ~~~~\left[\chi(\Omega) > \chi_{\rm tr}(\Omega)\right] \\ 
\end{array} 
\right. ,
\label{psn_all}
\end{equation}
where $d\Omega$ is the solid angle subtended by a neighboring cell, 
$E_{\rm SN} (=N_{\rm SN}\,10^{51}\,\rm{erg})$ is the total explosion energy, 
$M_{\rm ej}$ is the total ejecta mass, $n_{\rm H}$ and $Z$ are the hydrogen number density 
and gas metallicity of the cell into which the blast wave propagates, and 
$f_e = 1-\frac{\chi-1}{3(\chi_{\rm crit}-1)}$ is introduced to smoothly connect the two regimes.
The most important quantity in this equation is
\begin{equation}
\chi  \equiv dM_{\rm shell}(\Omega) /  dM_{\rm ej}(\Omega),
\end{equation}
which is the ratio of the shell (ejecta plus mass swept up) to ejecta mass along some solid angle ($\Omega, \Omega+d\Omega$) (see Equation~\ref{eq:mshell}).
The first part of Equation~\ref{psn_all} is valid up to the cooling phase and the second after that. 
A new and critical element is that we differentiate the energy-conserving and momentum-conserving phase 
by introducing a mass ratio at the transition ($\chi_{\rm tr}$). 
In actual implementations, the swept-up mass in direction $\Omega$ is taken as the sum of the 
gas mass in the adjacent cell in direction $\Omega$ and some fraction of the gas mass in the SN cell (see below).

A simple estimate of momentum budget per SN at the free expansion phase 
is $\sqrt{2 m_{\rm SNII} e_{\rm SN}} \approx 3.9\times10^4 \, {\rm km\, s^{-1}} \,\msun$ if the typical SN progenitor mass of 
$\bar{m}_{\rm SNII}=15.2\, \msun$ on the zero-age main sequence, appropriate for the Chabrier IMF \citep{chabrier03} 
with the lower (upper) mass of 8 (100) \msun, is used. 
On the other hand, the momentum of the SN bubble 
at the end of the adiabatic phase is much higher \citep{chevalier74,cioffi88,blondin98}, 
\begin{equation}
p_{\rm SN} (E,\,n_{\rm H}) \approx 3\times 10^5\,  {\rm km\, s^{-1}} \,\msun\, E_{51}^{16/17} n_{\rm H}^{-2/17}, 
\end{equation}
where $E_{51}$ is the energy in the unit of $10^{51} \,{\rm erg}$.
In addition, \citet{thornton98} showed that the momentum input is a decreasing 
function of the metallicity of ambient medium ($p_{\rm SN} \propto f(Z)\equiv \max\left[Z/Z_{\odot},0.01\right]^{-0.14}$). 
Below $0.01 Z_{\odot}$ atomic cooling is primarily responsible for radiative energy loss,
and no dependence with metallicity is seen. Motivated by this, we take the momentum 
\begin{equation}
p_{\rm SN} (E,\,n_{\rm H},\,Z) \approx 3\times 10^5\,  {\rm km\, s^{-1}} \,\msun\, E_{51}^{16/17} n_{\rm H}^{-2/17} f(Z), 
\label{psn_one}
\end{equation}
during the momentum-conserving phase (i.e., $\chi>\chi_{\rm tr}$). It is worth noting that the momentum transfer from SN explosions 
at the snowplow phase is essential to understanding the self-regulation of star formation in the local ISM \citep[e.g.][]{shetty12,kim13}.

\begin{figure}
   \centering
      \includegraphics[width=6.cm]{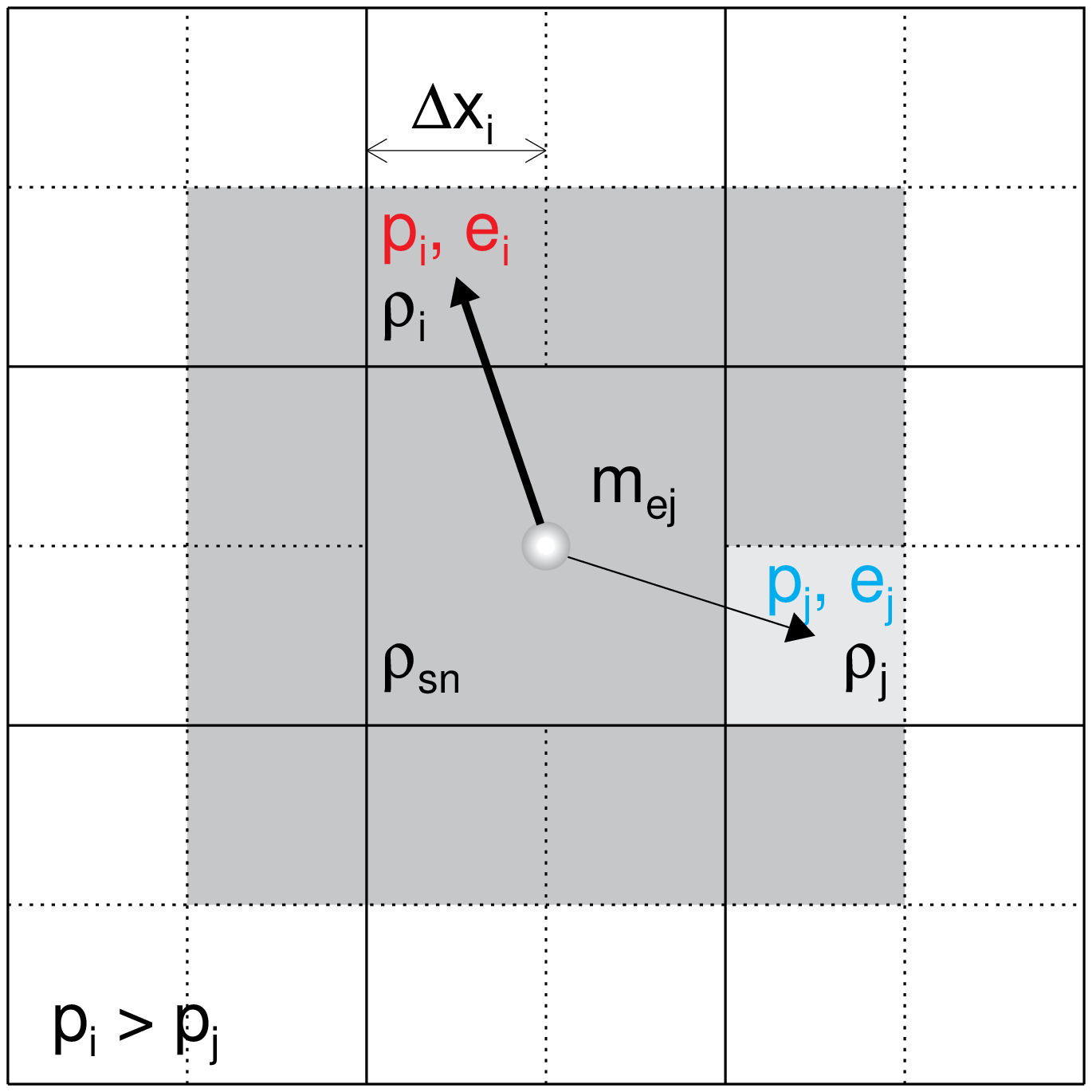}
      \caption{A schematic plot of a grid structure around a star particle that undergoes a SN explosion.
      Shaded regions display cells that are directly affected by the SN explosion. 
      The cell with the light grey color represents a low-density gas with $\chi<\chi_{\rm tr}$ 
      into which a smaller amount of momentum is deposited than other dense cells.
      }
      \label{sn_impl}
\end{figure}

The mass ratio at the transition ($\chi_{\rm tr}$) is estimated by equating Equation~\ref{psn_one} to 
$\sqrt{2\,\chi_{\rm tr} \,  M_{\rm ej} \, N_{\rm SN} e_{\rm SN,tr}}$, where 
$e_{\rm SN,tr}\approx 6.76\times10^{50} \, {\rm erg}$ is the kinetic energy at the transition \citep{blondin98}, 
and $N_{\rm SN}$ is the total number of SN events in a cell.
This yields
\begin{align}
\chi_{\rm tr} & = \frac{900}{0.676 \left(\bar{m}_{\rm SNII }/\msun\right)} \, E_{\rm 51}^{-2/17} n_0^{-4/17} \, f^2(Z)  \nonumber \\
                          & \simeq 87.54 \, E_{\rm 51}^{-2/17} n_0^{-4/17} \, Z'^{-0.28} , 
\end{align}
where $Z' = {\rm max} \left(Z/Z_{\odot}, 0.01\right)$.  Note that Equation~\ref{psn_all} is correct when energy is conserved,
and provides a good approximation until $\chi = \chi_{\rm tr}$.
On the other hand, all the momentum available ($p_{\rm SN}$) will be transferred to the surrounding gas 
when SNe explode in a cell in which a large amount of gas exists ($\chi > \chi_{\rm tr}$). 

Along with the momentum, mass (ejecta plus mass swept up) and energy are added to the neighboring cells.
To illustrate this, let us suppose that a cell in which a SN sits is surrounded by further refined cells (Figure~\ref{sn_impl}). 
In this case, the total number of adjacent cells except for the ones near 8 vertices is 48 in three-dimensional space. 
Note that even when the neighbors are not further refined, the following scheme can be applied 
pretending that the cells are composed of refined cells with the same physical properties.
We assume that the gas mass entrained from the SN cell and ejecta are evenly distributed (by the volume) 
to the cells that are directly affected by the SN (shaded region in Figure~\ref{sn_impl}).
The total shell mass entrained in a neighboring cell is then
\begin{equation}
dM_{\rm shell} = \rho_i \Delta x_i^3 + (1-\beta_{\rm sn}) \left( \rho_{\rm sn} \Delta x_{\rm sn}^3 + M_{\rm ej} \right) \frac{d\Omega}{4\pi} ,
\label{eq:mshell}
\end{equation}
where $\rho_{\rm sn}$ and $\Delta x_{\rm sn}$ are the density and size of the SN cell, respectively, 
$M_{\rm ej}=\alpha\, \sum m_{\star}$ is the ejecta mass from SN explosions in the SN cell,  and
$\alpha\simeq 0.317$ is the mass fraction of 8--100 \msun\ stars for a simple stellar population with the 
Chabrier IMF. The corresponding stellar ejecta in the neighboring cell is $dM_{\rm ejecta} =  (1-\beta_{\rm sn}) M_{\rm ej} d\Omega/4\pi$.
We take a simple approximation that $d\Omega/4\pi=1/48$. In Equation~\ref{eq:mshell}, 
$\beta_{\rm sn}$ determines how much fraction of the gas entrained is left behind 
in the SN cell. In the case of unigrid or smoothed particle hydrodynamics (SPH) simulations, this number can simply 
be chosen as the ratio of the volume taken by the SN cell and  total volume that are directly affected, i.e., 
$\beta_{\rm sn}=V_{\rm sn}/(V_{\rm sn}  +V_{\rm neighbor})$. 
We take $\beta_{\rm sn}=4/56$ so that mass is evenly distributed 
when the neighboring cell has the same level of refinement as the SN cell. 

Note that this scheme can easily be implemented in SPH simulations by computing the local mass 
loading through some solid angle for the N nearest particles.
While this paper was being written, we came to know that a similar approach for SN feedback 
is implemented by \citet{hopkins13}, based on the cooling radii of individual blast waves, 
$r_{\rm cool}\approx {\rm28\,pc}\, E_{51}^{0.29} n_0^{-0.43}\,f(Z)$.
It is useful to point out that there are two significant differences between the two models.
First, the momentum input at solar metallicity is about twice larger in \citet{hopkins13} than the measurements 
from a set of high-resolution hydrodynamics simulations of SN explosions that we use \citep{thornton98}.
Second, their input momentum has a steeper dependence on metallicity ($p_{\rm SN}\propto Z^{-0.27}$)  
than that our case  \citep[$p_{\rm SN}\propto Z^{-0.18}$,][]{thornton98}.
We note that the combination may result in the overestimation of the impact of SN explosions 
by a factor of $\sim$2.4 for the gas with $Z\approx0.01-0.1\,Z_{\odot}$ in their case.

Finally, we ensure that the total energy is conserved before and after the momentum injection; 
the residual (total minus kinetic) surplus energy is added as thermal energy to the affected cells in our case.

\begin{figure}
   \centering
      \includegraphics[width=7.5cm]{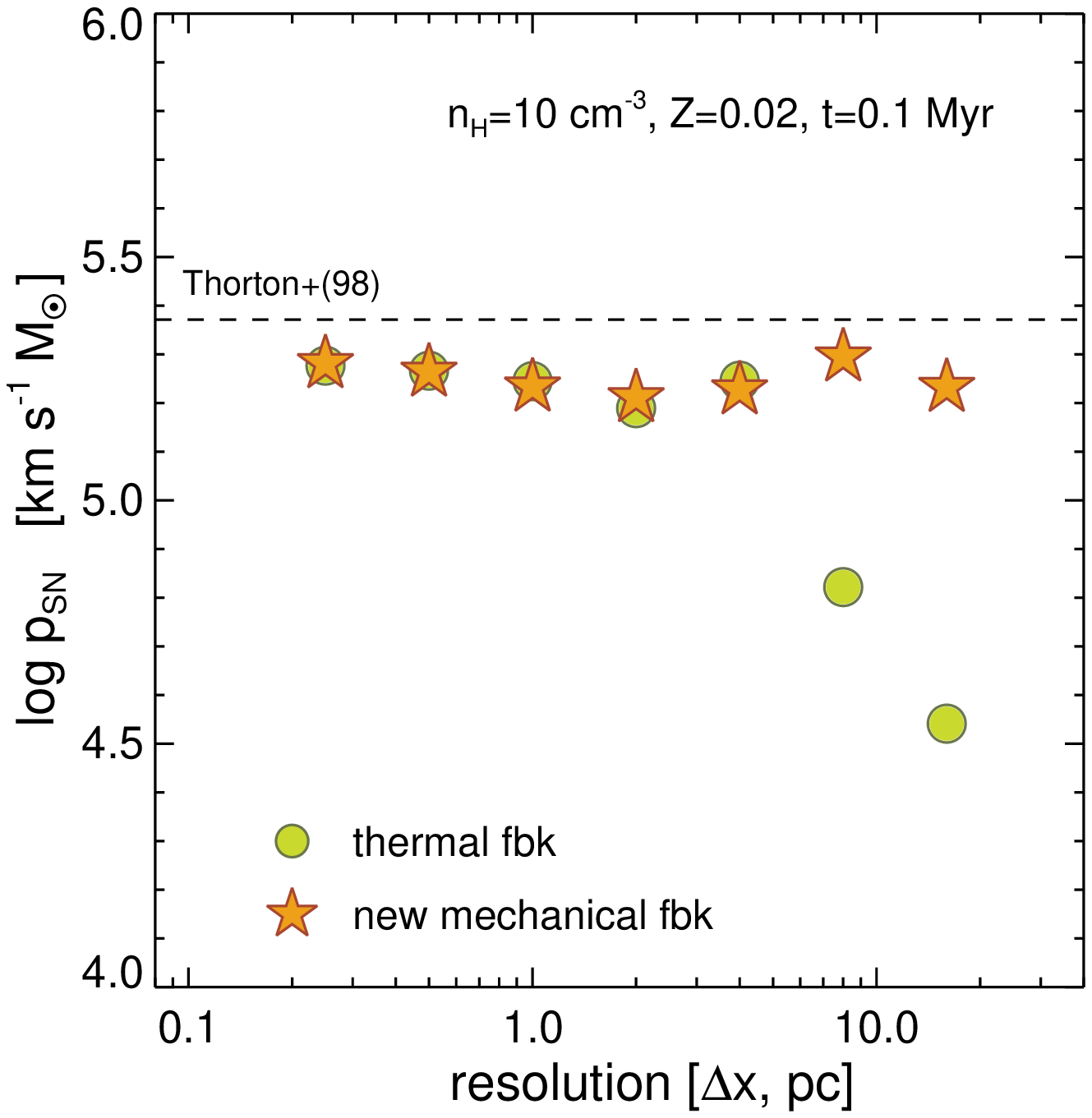}
      \caption{ Momentum transfer from a single SN event in a uniform medium of density 
      $n_{\rm H}=10\,{\rm cm^{-3}}$ and solar metallicity as a function of different 
      spatial resolution ($\Delta x_{\rm min}$). Note that the radiative cooling is included in this calculation.
      The radial momentum is measured at the momentum-conserving phase (0.1 Myr). 
      The dashed line shows the momentum in the shell predicted from the one dimensional hydrodynamic 
      calculation by \citet{thornton98}. Orange stars and green circles show the momentum transfer from the explosinon
      with our new model and thermal feedback often used in the literature, respectively.
      }
      \label{sn_vali}
\end{figure}

In order to examine how the new model compares with the standard energy feedback scheme, 
we perform idealized simulations by placing a SN in a uniform medium of number density 
$n_{\rm H}=10\,{\rm cm^{-3}}$ with solar metallicity. 
The radiative cooling is included in all calculations. The size of the simulated box is set to 128 pc 
(or 256 pc for the coarsest resolution run) while increasing the number of cells from $16^3$ to $512^3$. 
Then we measure the radial momentum from the explosion at the momentum-conserving phase \citep[0.1Myr,][]{thornton98}.
Note that the mass swept up by supernova ejecta is different depending on the resolution in our new feedback model, 
which in turn change the input momentum to the adjacent cells at the time of explosion.
For example, $\chi$ varies from 1.0006 to 164.3 for the runs with 0.25 to 16 pc, respectively. The corresponding  
momentum input based on Equation~\ref{psn_all} is 1.0003 to 7.136 times $\sqrt{2m_{\rm SNII} e_{\rm SN}}$,
and thus the different resolution runs correspond to different stages of the explosion from 
the adiabatic to momentum-conserving phase in practice.
 Figure~\ref{sn_vali} shows that approximately the same amount of momentum is transferred to the surrounding medium
with our new mechanical feedback scheme. For comparison, we also run the same set of simulations with 
thermal feedback which distributes the SN energy into the 27 surrounding cells.
We find that although the amount of momentum at the snowplow phase is roughly $20\%$ smaller 
than the prediction by 1D hydrodynamics simulations \citep{thornton98}, the results from the thermal 
and our mechanical model are very similar when high resolution is employed ($\Delta x \lesssim$ 4 pc).
On the other hands, the thermal feedback exhibits the well-known overcooling problem 
in lower resolution runs ($\Delta x > $ 8 pc).
We also perform the same experiment in gas with $n_{\rm H}=100\,{\rm cm^{-3}}$, and find that at least 
1pc resolution is required to properly model the momentum transfer with thermal feedback, 
while the momentum transfer is again not sensitive to the resolution with our mechanical feedback scheme.

\section{Suppression of star formation in low-mass halos}

In the main text, we show that stars form inefficiently in a low-mass halo 
with $10^8\lesssim \mvir \lesssim 10^9\,\msun$, and attribute this to the fact that supernova 
feedback is very effective. However, photoionizing background radiation may also prevent gas from collapsing 
into the small halos \citep[e.g.][]{shapiro94,thoul96}. In order to substantiate our claim, we carry out a 
hydrodynamics+N-body version of the \FRU\ run down to $z=9$, with a radiative transfer module turned off. 
Other physical and numerical parameters including the initial condition are kept fixed.
Note that we expect a stronger impact from reionization in the \FRU\ run, as the escape of ionizing photons 
is more significant than in the \FRR\ run.
Figure~\ref{fig:mz9} shows that only a small fraction of baryons ($<0.01\Omega_{\rm b}/\Omega_{\rm m}$) 
is converted into stars in the small halos even in the absence of ionizing radiation, 
demonstrating that supernova feedback is primarily responsible for the low conversion efficiency found in our simulations.

\begin{figure}
   \centering
      \includegraphics[width=7.cm]{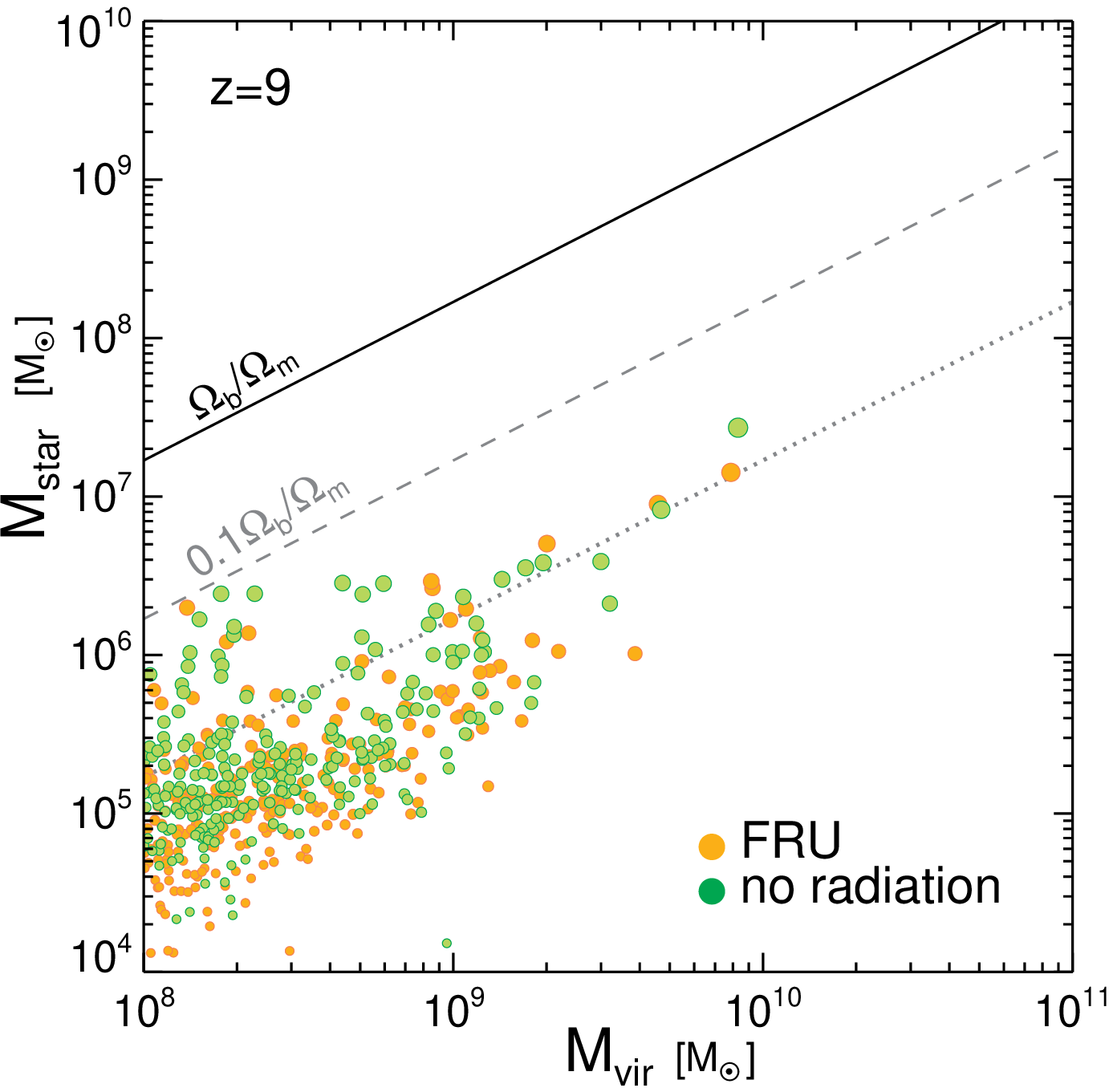}
      \caption{ Comparison of the galaxy stellar mass from the simulations with (orange) and 
      without (green) ionizing radiation at $z=9$. This plot shows that the low baryon-to-star conversion 
      efficiency of the central galaxy in our small mass halos ($\mvir\lesssim 10^9\,\msun$) is not 
      mainly due to the photoionizing background radiation, but due to supernova feedback.
      }
      \label{fig:mz9}
\end{figure}
}

\section{Effect of spatial resolution}

\begin{figure}
   \centering
      \includegraphics[width=7.cm]{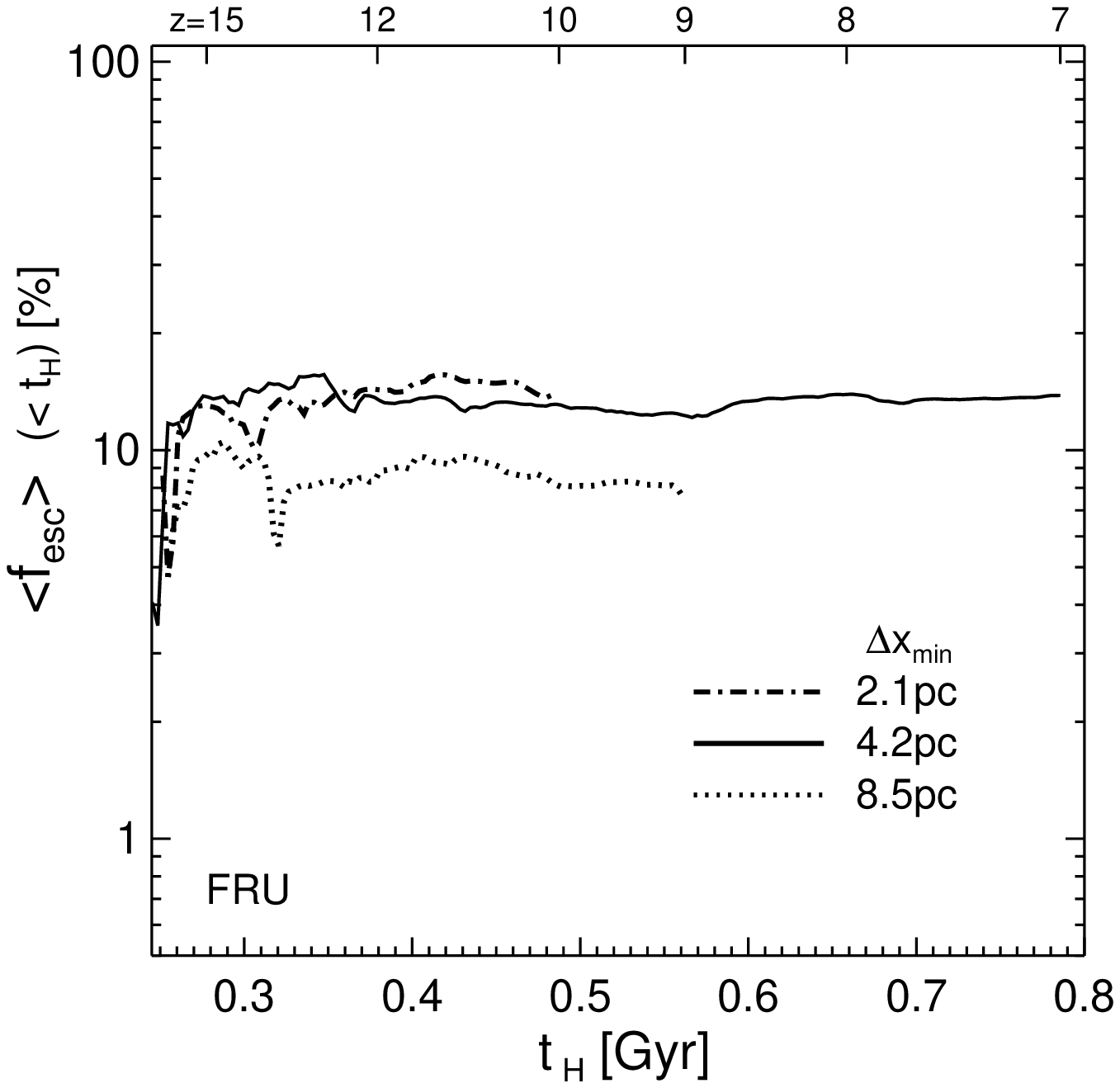}
      \caption{ Effect of spatial resolution on the photon number-weighted mean escape fraction.
      The solid line corresponds to $\fescg (< t_{\rm H})$ from the \FRU\ run, which employs the 
      maximum resolution of 4.2 pc (physical). The dotted and dot-dashed lines show the results with 
      one less or more level of refinement than the \FRU\ run, as indicated in the legend. 
      The mean escape fraction is reasonably converged in our fiducial run, 
      compared with the results from the higher resolution run.
      Lower resolution run shows a lower escape fraction due to less episodic star formation histories. 
      }
      \label{fig:res}
\end{figure}

We examine the effect of varying the spatial resolution on the mean escape fraction in Figure~\ref{fig:res}.
All other parameters are fixed as in the \FRU\ run. We run the lower (higher) resolution simulation down 
to z=9 (z=10), and compare the photon number-weighted time average of the escape fraction ($\fescg(<t_{\rm H})$) 
with that from the \FRU\ run. 
While the lower resolution run (8.5 pc, dotted) shows a lower mean escape of $\fescg=7.6\%$ than 
the \FRU, a similar fraction (13.9\%) of ionizing photons is escaped from halos in the higher resolution run (2.1pc, dot-dashed), indicating that the escape fraction in our fiducial run is reasonably converged for the parameters used.
We carefully inspect the cause of the lower escape fraction in the 8.5 pc run, 
and find that this is due to a longer time delay between the peak of star formation and escape fraction
as gas clouds collapse slowly compared with the higher resolution runs (2.1 or 4.2 pc).

\small
\bibliographystyle{apj}
\bibliography{ms}

\end{document}